\documentclass[5p,times,preprint]{elsarticle}

\usepackage{dingbat}
\usepackage[draft]{hyperref}
\usepackage{graphicx}
\usepackage{paralist}
\usepackage[english]{babel}
\usepackage{float}
\usepackage{makecell}
\usepackage{multirow}
\usepackage{amsfonts}
\usepackage{booktabs}
\usepackage{siunitx}
\usepackage{etoolbox}
\usepackage{subcaption}
\usepackage{multirow}
\usepackage{numprint}
\usepackage[T1]{fontenc}
\usepackage{algorithm}
\usepackage{flushend}
\usepackage[noend]{algpseudocode}
\usepackage{textcomp}
\usepackage{xcolor}
\usepackage{fancyvrb}
\usepackage{listings}
\usepackage{paralist}
\usepackage[inline]{enumitem}
\usepackage{booktabs}
\usepackage{color}
\usepackage{colortbl}
\usepackage{listings}
\usepackage{flexisym}
\usepackage{array}
\usepackage{xparse}
\usepackage{xspace}
\usepackage[flushleft]{threeparttable}
\usepackage[nobreak]{mdframed}
\usepackage{framed}

\algnewcommand{\Inputs}[1]{%
  \Statex \textbf{Inputs:}
  \Statex \hspace*{\algorithmicindent}\parbox[t]{.8\linewidth}{\raggedright #1}
}
\algnewcommand{\Outputs}[1]{%
  \Statex \textbf{Outputs:}
  \Statex \hspace*{\algorithmicindent}\parbox[t]{.8\linewidth}{\raggedright #1}
}
\algnewcommand{\Initialize}[1]{%
  \State \textbf{Initialize:}
  \Statex \hspace*{\algorithmicindent}\parbox[t]{.8\linewidth}{\raggedright #1}
}

\def\cca#1{\cellcolor{black!#1}\ifnum #1>49\color{white}\fi{#1}}

\makeatletter
\NewDocumentCommand{\LeftComment}{s m}{%
  \Statex \IfBooleanF{#1}{\hspace*{\ALG@thistlm}}\(\triangleright\) #2}
\makeatother
 
\definecolor{dkgreen}{rgb}{0,0.6,0}
\definecolor{gray}{rgb}{0.5,0.5,0.5}
\definecolor{mauve}{rgb}{0.58,0,0.82}
\definecolor{gray}{rgb}{0.4,0.4,0.4}
\definecolor{darkblue}{rgb}{0.0,0.0,0.6}
\definecolor{lightblue}{rgb}{0.0,0.0,0.9}
\definecolor{cyan}{rgb}{0.0,0.6,0.6}
\definecolor{darkred}{rgb}{0.6,0.0,0.0}

\lstdefinelanguage{XML}
{
  morestring=[s][\color{mauve}]{"}{"},
  morestring=[s][\color{black}]{>}{<},
  morecomment=[s]{<?}{?>},
  morecomment=[s][\color{dkgreen}]{<!--}{-->},
  stringstyle=\color{black},
  identifierstyle=\color{lightblue},
  keywordstyle=\color{red},
  morekeywords={xmlns,xsi,noNamespaceSchemaLocation,type,id,x,y,source,target,version,tool,transRef,roleRef,objective,eventually,name,id,type,role,entityId,distance,parentId}% list your attributes here
}

\definecolor{javared}{rgb}{0.6,0,0} % for strings
\definecolor{javagreen}{rgb}{0.25,0.5,0.35} % comments
\definecolor{javapurple}{rgb}{0.5,0,0.35} % keywords
\definecolor{javadocblue}{rgb}{0.25,0.35,0.75} % javadoc

\lstdefinestyle{diff}{
    escapechar=\%
}

\usepackage[]{xcolor}
\newcommand{\TODO}[1]{\textcolor{red}{#1}\GenericWarning{}{LaTeX Warning: TODO: #1}}\newcommand\todo\TODO

\newcommand{\revision}[1]{\textcolor{black}{#1}}
\newcommand{\revisiontwo}[1]{\textcolor{black}{#1}}
\newcommand{\projectsFromBears}{\textsc{PB}\xspace}

\newcommand{\groundtruthpatches}{\textsc{Ground-truth}\xspace}

\newcommand{\toolname}{\textsc{LighteR}\xspace}

\newcommand{\repairspacecommit}{{repair-space commit}\xspace}
\newcommand{\repairspacecommits}{{repair-space commits}\xspace}
\newcommand{\GHilight}{\makebox[0pt][l]{\color{green}\rule[-4pt]{1\linewidth}{10pt}}}
\newcommand{\RHilight}{\makebox[0pt][l]{\color{pink}\rule[-4pt]{1\linewidth}{10pt}}}

\def\BibTeX{{\rm B\kern-.05em{\sc i\kern-.025em b}\kern-.08em
    T\kern-.1667em\lower.7ex\hbox{E}\kern-.125emX}}

\begin{document}

\begin{frontmatter}

\title{Estimating the Potential of Program Repair Search Spaces with Commit Analysis}
%\title{Fast and Lightweight Analysis of the Potential of Program Repair Approaches}


\author[kth]{Khashayar Etemadi\corref{mycorrespondingauthor}}
\cortext[mycorrespondingauthor]{Corresponding authors}
\ead{khaes@kth.se}

\author[sharif]{Niloofar Tarighat}
\ead{tarighat\_niloofar@ee.sharif.edu}

\author[india]{Siddharth Yadav}
\ead{siddharth16268@iiitd.ac.in}

\author[france]{Matias Martinez}
\ead{matias.martinez@uphf.fr}

\author[kth]{Martin Monperrus}
\ead{monperrus@kth.se}

\address[kth]{KTH Royal Institute of Technology, Sweden}
\address[sharif]{Sharif University of Technology, Iran}
\address[india]{Indraprastha Institute of Information Technology Delhi, India}
\address[france]{Universit\'e  Polytechnique Hauts-de-France, France}

\begin{abstract}
The most natural method for evaluating program repair systems is to run them on bug datasets, such as Defects4J.
Yet, using this evaluation technique on arbitrary real-world programs requires heavy configuration.
\revision{In this paper, we propose a purely static method to evaluate the potential of the search space of repair approaches.  \revisiontwo{This new method enables researchers and practitioners to encode the search spaces of repair approaches and select potentially useful ones without struggling with tool configuration and execution.}}
We encode the search spaces by specifying the repair strategies they employ. Next, we use the specifications to check whether past commits lie in repair search spaces. For a repair approach, including many human-written past commits in its search space indicates its potential to generate useful patches.
We implement our evaluation method in \toolname. \toolname gets a Git repository and outputs a list of commits whose source code changes lie in repair search spaces.
We run \toolname on 55,309 commits from the history of 72 Github repositories with and show that \toolname's precision and recall are 77\% and 92\%, respectively.
\revision{Overall, our experiments show that our novel method is both lightweight and effective to study the search space of program repair approaches.}
\end{abstract}


\end{frontmatter}

\section{Introduction}

%%% APR + dataset
Fixing software bugs is a notoriously time-consuming task for developers \cite{Murphy-HillZBN13}.
To address this issue, automatic program repair (APR) approaches apply repair strategies to fix software bugs without human intervention \cite{Monperrus2015}. 
Researchers usually assess repair approaches by running them on bug datasets, such as Defects4J \cite{just2014defects4j}, Bugs.jar \cite{saha2018bugs} and ManyBugs \cite{GouesBFW17}. Comparative evaluations of repair systems (e.g., \cite{liu2020efficiency, martinez2017automatic, durieux2019empirical}) have shown promising results in terms of the number of bugs  that can be fixed in a given dataset.

%%% obstacles of fast evaluation
Even though executing repair approaches is the most natural method for evaluating APR, there are two main obstacles when this evaluation is done on an arbitrary software project. First, fully executing a repair approach on a real world project often requires heavy and time-consuming configuration of the repair approach and the target project \cite{durieux2019empirical}. 
Second, the target programs under repair should have a test suite specifying their correct behaviour and, at least, one failing test case that exposes the bug. \revision{Previous work~\cite{tufano2017there,madeiral2019bears} showed it is hard to find real world commits with test suites that can be compiled and executed. These two major obstacles (configuration and dependability on strong testing) hinder assessment of automated program repair on new projects.}

%%% our method, search space, commit coverage
\revision{\revision{In this paper, we propose a new lightweight method to check whether repair approaches may be fruitful for new projects.}} Instead of fully executing a repair system to check if it actually fixes certain bugs, we analyze whether real world bug-fixing commits lie in the considered repair search space \cite{martinez2015mining}. 
In this context, the search space of the repair approach is the set of all program patches that can be potentially generated. 
For example, GenProg~\cite{le2011genprog}'s search space contains all replacements of statements with a new one, where the new statement is copied from the program under repair.
In this work, we first specify search spaces of repair approaches based on code patterns. 
This enables us to then compare real world commits against our specifications of a repair approach search space as follows. 
For each commit, we check if it lies in the search space of a known repair approach.
If that happens, we say that the approach covers that commit, and vice versa, that the commit could have been potentially produced by the repair approach. 
\revision{A repair approach that covers many real world human-written commits is likely to create useful patches in the future. In other words, the higher the commit coverage, the broader the search space of the approach.}

%%% the tool and experiments
We implement our novel method in \toolname. For a given commit that changes a program, \toolname performs static analysis on the abstract syntax tree (AST) to determine if the commit is covered by certain repair approaches. 
\toolname specifies the search spaces of $8$ notable repair systems: Arja~\cite{yuan2018arja}, Cardumen~\cite{martinez2018ultra}, Elixir~\cite{saha2017elixir}, GenProg~\cite{le2011genprog}, jMutRepair~\cite{martinez2016astor}, Kali~\cite{qi2015analysis}, Nopol~\cite{xuan2016nopol}, and NPEfix~\cite{cornu2015npefix}. 

%%% experiments and results
We run \toolname on $55,309$ commits of $72$ projects from \textsc{Bears}~\cite{madeiral2019bears} to study the effectiveness of \toolname.
Our experiments show that $747/55,309$ of all the considered commits lie in the search space of at least one repair system.
We also demonstrate that there is little overlap between the repair systems, showing that program repair research is producing systems that are complementary in practice.
The median time \toolname spends to check a commit against the search space of a repair approach is $0.81$ second, which is fast enough for practical usage.
In another experiment, we measure how accurately \toolname determines whether a commit lies in the search space of a repair system compared to a ground-truth classification.
Our results show that \toolname has a precision and recall of $77\%$ and $92\%$ respectively.
\revision{Overall \toolname is useful to estimate the potential of the search space of repair approaches on a new project, without going through full configuration and execution of actual tools.}

% RW/ novelty statement
\revision{Few studies have analyzed the search spaces of repair approaches \cite{martinez2015mining,long2016space}, and all of them with a different purpose than ours. We are the first to propose a lightweight method for conducting a fast evaluation of the breadth of the search space of repair approaches}.

To sum up, our contributions are:
\begin{itemize}
    \item A novel method for specifying the search space of program repair approaches, appropriate to study the potential of program repair to create patches corresponding to past commits of software repositories. This framework implemented in a tool called \toolname, is lightweight, it does not require configuration and execution of repair systems.
    \item A comprehensive series of experiments on past commits. By analyzing $55,309$ human-written commits from $72$ Github repositories, we show that $1.35\% \, (747/55,309)$ of past commits lie in the search space of at least one of the considered repair systems, $62\%$ of these commits are indeed bug-fixing commits according to the  manual inspection we conducted. 
    Overall, our experiments show that our novel method is an effective means to study the potential of program repair search space.
    \item A systematic measurement of the reliability of \toolname. Our prototype system has a precision and recall of $77\%$ and $92\%$, respectively, which is arguably high compared to close tools, such as PPD~\cite{madeiral:hal-01851813}.
\end{itemize}

The rest of this paper is organized as follows: 
\autoref{sec:terminology} presents the terminology that we use in this paper.
\revision{\autoref{sec:eval_types} presents the different types of evaluation in program repair.} \autoref{sec:approach} describes our proposed method. \autoref{sec:methodology} and \autoref{sec:results} explain the methodology and then the results of our experiments. Implications of our results are discussed in \autoref{sec:discussion}. \autoref{sec:related} reviews the related work. Finally, \autoref{sec:conclusion} concludes this paper.

\section{Terminology}
\label{sec:terminology}
We use the following concepts throughout this study:

\noindent\textbf{Automatic Repair Approach:} A software artifact that gets a buggy version of a program as input and generates patches that fix the bug as output \cite{weimer2009automatically}. To generate the patches, a repair approach also requires an oracle that determines whether a version of a program is buggy or correct. For example, \emph{test-suite based program repair approaches} use test-suites as the oracle \cite{le2011genprog}.

\noindent\textbf{Repair Operator:} A type of atomic change that is applied on the buggy program to repair the bug. For example, removing a statement from the source code is an operator used by Kali~\cite{qi2015analysis}.
    
\noindent\textbf{Repair Strategy:} A set of repair operators applied in conjunction by a repair approach to the buggy version of a program.
For example, one of the strategies employed by NPEfix \cite{cornu2015npefix} is ``skip method'' (e.g., see \autoref{lst:npefix_ex}). Per this strategy, an if-statement is added before a suspicious statement. The corresponding if-condition checks whether a variable used by the suspicious statement is equal to null. If the if-condition holds, a return statement is executed.

\begin{lstlisting}[float=t, style=diff, caption={\revision{Commit 275c6fdb in Jgrapht, which is in the search space of NPEFix, it applies repair strategy ``skip method''.}}, captionpos=b, label=lst:npefix_ex]
  Set<E> removed = getAllEdges(sourceVertex, targetVertex);
%\GHilight%+ if(removed == null) {
%\GHilight%+     return null;
%\GHilight%+ }
  removeAllEdges(removed);
\end{lstlisting}
%\end{pyglist}
    
\noindent\textbf{Repair Ingredient:} An existing source code fragment that is reused by a repair approach to fix the bug \cite{martinez2014fix,white2019sorting}. For example, in one of its repair strategies, GenProg \cite{le2011genprog}  creates a candidate patch by replacing a suspicious statement by another existing statement written elsewhere in the program.
The latter is the \emph{ingredient} of the candidate patch.
Note that ingredients can have different granularities.
For example, in GenProg, an ingredient is a statement, in NPEfix \cite{cornu2015npefix} it is a variable, and in Cardumen \cite{martinez2018ultra} it is an expression.

\noindent\textbf{Scope of Ingredients:} The scope of ingredients is the parts of program that are considered for extracting repair ingredients \cite{martinez2014fix, white2019sorting}. 
For example, jGenProg \cite{martinez2016astor} can replace an old statement $s$ (written in file $f$ from package $p$) with a new one, according to three different scopes:
\begin{inparaenum}[\it 1)]
\item same file (i.e., $f$),
\item same package (i.e., from any file belonging to $p$), and
\item same program.
\end{inparaenum}

%the same file, package, or program.
%This means that the scope of ingredients for JGenProg can be ``same file'', ``same package'', or ``same program''.
    
\noindent\textbf{Search Space of Repair Approach:} Let us assume a repair approach $r$ with certain repair strategies and a scope of ingredients. When a program is given as the input, the search space of $r$ is the set of all patches that $r$ can generate given the strategies and scope of ingredients \cite{martinez2015mining}.

\section{Types of Evaluation in Program Repair}
\label{sec:eval_types}

\revision{There are various ways for evaluating program repair approaches. In this section, we classify these techniques into two categories, dynamic evaluation and static evaluation, and we discuss their use cases and limitations.}

\subsection{\revision{Program Repair Steps Considered in Scientific Evaluation}}

\revision{\textbf{Assumption Verification:} Test-based repair approaches assume the presence of a failing test that exposes the bug. Similarly, the repair system should be able to successfully build the program under repair before fixing it. These are core assumptions of test-suite based repair. Bug datasets facilitate program repair research by curating the buggy programs that meet those repair assumptions \cite{madeiral2019bears}. We note that many bugs and their fixes exist in repositories without satisfying those assumptions, yet providing valuable knowledge for program repair research.}

\revision{\textbf{Fault Localization:} This step refers to the process of ranking locations in the buggy program based on their likelihood to cause a bug \cite{wong2016survey}. Repair approaches take advantage of fault localization methods to find the best candidate locations that should be changed to fix a bug. It is possible to isolate fault localization in program repair to study its importance \cite{liu2019you}.}

\revision{\textbf{Ingredient Extraction:} Redundancy based program repair approaches have an ``ingredient extraction'' step. In this step, the repair system extracts code components in the existing program that may be used for patch generation. This step is usually performed statically. Researchers have studied this step in isolation \cite{martinez2014fix,Barr2014Surgery}.}

\revision{\textbf{Code Synthesis:} A program repair patch is composed of code that is synthesized, possibly from ingredients in the case of redundancy based repair \cite{le2011genprog}. To do this, templates and code transformations are applied on the AST of the program. This step is static in most of the related work, with the exception of the dynamic collection of ingredients in \cite{durieux2016dynamoth}. It results in a set of candidate patches. An example of a study of code synthesis which is purely static is by Martinez et al. \cite{martinez2015mining}}

\revision{\textbf{Test Validation:} \revisiontwo{This is the step where all the tests are executed on the candidate patches, in order to discard the incorrect patches that do not pass them.} An example of a study dedicated to this step is \cite{le2018overfitting}}.

\revision{\textbf{Overfitting Detection:} The patches that pass all the tests but introduce regressions are filtered out based on static \cite{YeODS} or dynamic analysis \cite{xiong2018identifying}. This is called the overfitting detection step. This has been studied in isolation for example in \cite{long2016space}.}

\revision{\autoref{tab:pipeline_eval} summarizes those different steps of program repair. The ``Kind'' column shows if the corresponding step of repair is carried out statically or if it requires execution of the program under repair. The ``Focus'' column shows an example of studies that specifically evaluate a given step. Finally, the last column indicates whether the corresponding step is considered in our novel evaluation technique proposed in this paper.}

\subsection{\revision{Types of Evaluations}}

\begin{table}[tb]
\centering
\small
\caption{\revision{Execution steps of program repair approaches and example studies focus on them in isolation.}}
\label{tab:pipeline_eval}
\begin{tabular}{l l l c}
\toprule
	\textbf{Step} & \textbf{Kind} & \textbf{Focus} & \textbf{This Paper} \\ \hline
	\midrule
	Assumption Verification & Static & \makecell{\cite{madeiral2019bears}} &\\ \hline
	Fault Localization & Dynamic & \makecell{ \cite{liu2019you}}& \\ \hline
	Ingredient Extraction & Static &
	\makecell{\cite{martinez2014fix,Barr2014Surgery}} & \checkmark\\ \hline
	Code Synthesis & Static & \makecell{\cite{martinez2015mining}} & \checkmark \\ \hline
	Test Validation & Dynamic & \makecell{\cite{le2018overfitting}} &\\ \hline
	Overfitting Detection & \makecell{Dynamic/ \\ Static} & \makecell{\cite{long2016space,Le2016HDRepair}} &\\
	\bottomrule
\end{tabular}
\end{table}

\revision{Now, it is clear that we classify the evaluations in program repair into two main groups: dynamic evaluation and static evaluation.}

\revision{\textbf{Dynamic evaluation} methods focus on the dynamic steps of the repair process, and typically consist of running actual repair tools. By running the actual repair tools, these evaluation techniques may produce actual patches generated by the tools.
For example, the RepairThemAll study \cite{durieux2019empirical} executed 11 repair tools over five benchmarks, this is a archetypal dynamic evaluation. Evaluations of this type heavily depend on the feasibility of execution. For this reason, they require fine-tuned bug datasets appropriate for running repair tools on them, such as Defects4J \cite{just2014defects4j}. Because of this big curation effort, only a few datasets have been created accordingly. This leads to repair tools over-engineered to fix specific bugs in those datasets, which in turn causes overestimation of the generalizability \cite{durieux2019empirical}. The main advantage of dynamic evaluations is that it give concrete insights for practitioners. The main limitation is that it is very costly, and thus tends to be limited to the same bugs or benchmarks again and again.
}

% nothing about lighter here
\revision{\textbf{Static evaluation} methods focus on static studying some steps of the repair process, namely, ingredient extraction, code synthesis, and static overfitting detection. Static evaluation does not require collecting dynamic execution data for repair approaches, hence it can be performed without running the actual repair tools. For example, Martinez and Monperrus \cite{martinez2015mining} study the frequency of applying repair operators in a large dataset of human-made commits. This means they only consider the code synthesis step, which can be evaluated by a static analysis of target commits. The main advantage of static evaluation is that it saves engineering resources and experimental time (for environment setup, configuration, and execution) and it does not require presence of extensive test suites.
The main limitation is its abstractness, it does not tell which candidate patches will be delivered by actual repair tools.}

% paragraph about lighter
\revision{In this paper, we fully concentrate on repair strategies and ingredients. Thus, it fits perfectly with static evaluation. We  propose a conceptual framework and its implementation in \linebreak \toolname to statically evaluate the search space of repair approaches and their strategies. \toolname fully focuses on the ingredient extraction and the code synthesis step of the repair process, which are both amenable to static study. In other words, we statically investigate the potential of repair approaches in terms of the breadth of their search spaces.}.

\section{A Lightweight Method for Analyzing Program Repair Search Spaces}

\label{sec:approach}
\subsection{Overview}
The goal of our proposed method is to evaluate and compare different repair approaches in terms of the number of human-written patches that lie in their search space. 
For this purpose, we specify the search spaces of well-known repair approaches, and analyze the past commits of open-source repositories to compute the \emph{commit coverage} of each repair approach, as follows. To define commit coverage, we first define \repairspacecommit.

\noindent\textbf{Repair-space Commit:} Given a commit $c$ that transforms the \emph{old\_version} of a program into its \emph{new\_version} and a repair approach $r$, we say that commit $c=(old\_version, new\_version)$ is a \emph{\repairspacecommit} for $r$ if and only if \emph{new\_version} is in the search space of $r$ when \emph{old\_version} is given as the input.

Consider \autoref{lst:npefix_ex}, which is a real commit in project \linebreak Jgrapht\footnote{\revision{https://github.com/jgrapht/jgrapht/commit/275c6fdb}.}. This commit is an example of \repairspacecommits of the repair search space of NPEfix \cite{cornu2015npefix}. The commit contains a typical NPEfix null check characteristic of its search space. Hence we say that this commit is a \repairspacecommit for NPEfix. In this example, line 1 and line 5 are from the old\_version, while in the new\_version lines 2-4 are added. According to the definition of NPEFix, new\_version is in its search space because NPEfix has a ``skip method'' strategy that is able to produce this patch.

\noindent\textbf{Commit Coverage (CC):} The \emph{commit coverage} of repair approach $r$ over a set of commits $S$ is the number of commits in $S$ that are \repairspacecommits for $r$ divided by the total number of commits in $S$.

We consider real human-written patches to be useful \linebreak patches. Therefore, if a repair approach has a high commit coverage over a large dataset of human-written patches, this indicates the potential usefulness of the actual implementation of the repair approach.

In this section, we design a framework to detect \repairspacecommits. When the \repairspacecommits are detected, computing the commit coverages is trivial.
\revision{In this framework, we encode each repair strategy by specifying
\begin{inparaenum}[\it a) ]
\item the repair operators from the strategy, expressed using fine-grained code changes, and
\item the rules that those changes must respect (e.g., the code introduced by a change is a valid ingredient according to a given scope).
\end{inparaenum}}

\begin{figure*}
\centering
% source is file XXXX
\includegraphics[width=4.2in]{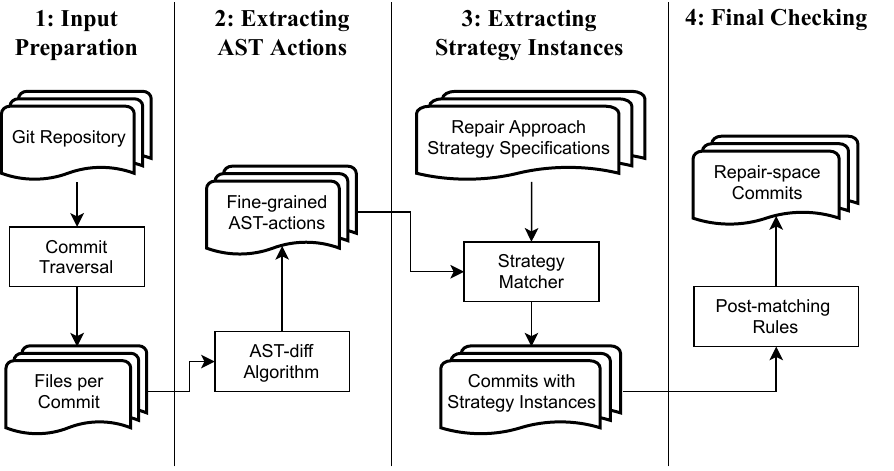}
\caption{Overview of the approach.}
\label{fig:coming}
\end{figure*}

% paragraph presenting the figure
\autoref{fig:coming} shows an overview of how the proposed approach \ works.
The whole process consists of four steps.
\begin{inparaenum}[\it 1)]
    \item Input preparation: for a given Git repository,  we identify the updated files for each commit  (see \autoref{sec:input}).
    \item Extracting AST actions: for each updated file, the actions that transform the AST of the old version into the new one are extracted  (see \autoref{sec:extracting_ast_actions}).
    \item Extracting strategy instances: the updated files whose corresponding AST actions match a ``strategy specification'' (which is defined in \autoref{sec:change_pattern_specification}) are determined (see \autoref{sec:extracting_instances}). We design the strategy specifications to model the repair strategies employed by the considered repair approaches.
   \item Final check: the commits whose updated files match strategy specifications are checked for additional constraints (see \autoref{sec:post_matching}). The result of this last step are the detected \repairspacecommits.
\end{inparaenum}

\subsection{Challenges}
\label{sec:challenges}

\begin{lstlisting}[float=t, style=diff, caption={A code change which could be seen as a ``statement replacement'', ``expression replacement'', or ``operator removal''.}, captionpos=b, label=lst:replace_statement_ex]
  if (var == val1){
    var = val1 + val2;
  } else if (var == val2) {
    var = val1;
  } else {
%\RHilight%-   var = val1 + val2;
%\GHilight%+   var = val1;
  }
\end{lstlisting}

The major challenge of repair-space commit detection is to find a representation of AST modifications that is appropriate for capturing program repair strategies.
For example, consider \autoref{lst:replace_statement_ex}. The AST modifications in this example can be represented in many different ways incl.:
\begin{inparaenum}[\it 1)]
    \item R1: It can be seen as a ``statement replacement'' action: a statement (line 6) is replaced by a new statement (line 7). The new statement is copied from line 4 of the same file.
    \item R2: It can be seen as an ``expression replacement'' action: an expression (``val1+val2'') from line 6 is replaced by a new expression (``val1''). The new expression is copied from line 4 of the same file.
    \item R3: It can be seen as an ``operator removal'' action: an operator and the corresponding operand (``+ val2'') is removed from line 6 and the statement at line 7 is the result.
\end{inparaenum}

If the code change at~\autoref{lst:replace_statement_ex} is represented as R1 (i.e., replace the statement in line 6 by another one), it lies in the search space of Arja and GenProg because those approaches are able to generate patches that replace one buggy statement by another. 
If it is represented with R2 (i.e., replacement of an expression by another one), it lies in the search space of Cardumen because Cardumen repairs bugs by replacing expressions. 
Finally, if it is represented with the third option, it does not lie in the search space of any of these three repair approaches because none of them has a repair operator that consists in removing a `+` operator. To sum up, one of the main research challenges that we are addressing in this study is to find the right AST action representation, which is appropriate for specifying repair search spaces.

\subsection{Input Preparation}
\label{sec:input}

\begin{algorithm}[!tb]
\caption{Algorithm of the proposed approach.}\label{tool_algorithm}
\begin{algorithmic}[1]
\Inputs{\textit{git\_repo:} The given Git repository \\ \textit{repair\_approach:} The repair approach whose search space should be considered}
\Outputs{\textit{repair\_space\_commits:} The set of detected \repairspacecommits for \textit{repair\_approach}}
\State $commits \gets get\_commits(git\_repo)$
\State $specs \gets get\_specifications(repair\_approach)$ \label{algo:get_specs}

\For {each commit $c$ in $commits$}
    \State $SI \gets []$ /* SI: strategy\_instances */
    %\Comment{SI: strategyInstances}
    \If{$only\_one\_file\_updated(c)$} \label{algo:checkUpdatedFilesCnt}
       \State  $f \gets get\_updated\_file(c)$ \label{algo:getUpdatedFile}
       \State  $f_p \gets get\_previous\_version(f)$ \label{algo:gerPrevious}
       \State  $f_n \gets get\_new\_version(f)$ \label{algo:getNew}
        \State $ES \gets GetDiff(f_p, f_n)$ /* ES: Edit Script */ \label{algo:computediff}
        %/* CO: changeOperations */
        \For{each specification $s$ in $specs$}
            \If{$match(ES, s)$} \label{algo:matching}
                \State $SI.insert(ES)$
            \EndIf
        \EndFor
    \EndIf
    %%to find a more elegant way
    \If{$pass\_post\_rules(SI, c, repair\_approach)$} \label{algo:post_checking}
        \State $repair\_space\_commits.insert(c)$
    \EndIf
    
\EndFor
\end{algorithmic}
\end{algorithm}

\autoref{tool_algorithm} shows how our technique works.
It takes as input the path to a Git repository and a repair approach whose search space should be considered. 
Then, it traverses over the history of the repository from the oldest commit to the most recent one. 

\revision{For each commit $c$, \toolname checks that only one file is updated (line \ref{algo:checkUpdatedFilesCnt}). Changes in multiple updated files cannot be covered by a single strategy instance, while we only target single-instance fixes that lie in the repair search space (see \linebreak \autoref{sec:pattern_matching}). Therefore, we discard commits with multiple updated files.
If $c$ has one updated file, \toolname gets that file $f$ (line \ref{algo:getUpdatedFile}) and constructs a pair of files \textit{<$f_p$, $f_n$>}, where <$f_p$> is the version of $f$ previous to $c$ (retrieved in line \ref{algo:gerPrevious}), and  <$f_n$> is the new version obtained after $c$ (retrieved in line \ref{algo:getNew}).}

\subsection{Extracting AST Actions from Updated Files}
\label{sec:extracting_ast_actions}
In the second step, \toolname computes the AST differences between the pair of files <$f_p$, $f_n$> (line \ref{algo:computediff}).
The output of this step is an \emph{edit script (ES)}, a list of actions that transforms $f_p$ into $f_n$.

\toolname uses GumTree algorithm~\cite{falleri2014fine} to compute these actions at the AST level.
In GumTree, there are four types of action: 
\begin{inparaenum}[\it 1)]
\item update, which changes the value of an AST node,
\item insert, which inserts a new AST node,
\item delete, which deletes an existing AST node, and
\item move, which moves an AST node and makes it child of another node.
\end{inparaenum}
These extracted fine-grained AST actions are passed to the next step.

\subsection{Extracting Strategy Instances}
\label{sec:extracting_instances}

This step determines if the fine-grained AST actions from an edit script correspond to those that can be synthesized by a repair approach. 

For each repair approach, we come up with one or more strategy specifications (described in \autoref{sec:change_pattern_specification}) that define its search space. Strategy specifications are abstract representations of the repair strategies employed by repair approaches.

% we refine the challenge in the scope of Gumtree
If AST actions in an edit script $ES$ match with a strategy specification $s$, we say that the $ES$ is an \emph{instance} of the $s$.

We now describe the specification language and then the matching process.

\subsubsection{Strategy Specification}
\label{sec:change_pattern_specification}

\begin{lstlisting}[language=XML, caption={One of the strategy specifications for jMutRepair.}, captionpos=b, label=jmutrepair_pattern, float=bt, floatplacement=bt, escapechar=|]
<pattern name="binary_upd">
  <entity id="1" type="If"/>|\label{pat:ent1}|
  <entity id="2" type="BinaryOperator">|\label{pat:ent2}|
    <parent parentId="1">|\label{pat:parent}|
  </entity>
  <action entityId="2" type="UPD"/>|\label{pat:action}|
</pattern>
\end{lstlisting}

Each specification uses an abstract representation to specify a certain repair strategy of a program repair approach.
The specifications are represented in the \textit{change pattern specification language} of \cite{martinez2019coming}, which we now briefly present. A strategy specification consists of a set of \textit{actions}, and each action is performed on an \textit{entity}. The types of actions of specifications are the same as the types of AST actions that GumTree extracts (\textit{update}, \textit{insert}, \textit{delete}, and \textit{move}).
In addition to these action types, a strategy specification can also contain an action of type \textit{unchanged}, which indicates an entity should not be changed (i.e., not affected by any action). Finally, a strategy specification can also define parenthood relations between entities.

For example, \autoref{jmutrepair_pattern} is a specification that corresponds to a repair strategy used by jMutRepair \cite{martinez2016astor}. According to this strategy, a binary operator inside an if-condition can be changed to another operator. Line~\autoref{pat:action} of the \autoref{jmutrepair_pattern} represents the update action. As it is stated, the ``entityId'' of the subject entity is ``2''. Therefore, this action is performed on the entity defined in line~\autoref{pat:ent2}. As shown in the specification, the type of this entity is ``BinaryOperator'' and the id of its parent is ``1'' (see line~\autoref{pat:parent}). Finally, the parent entity is defined in line~\autoref{pat:ent1} and as it is mentioned there its type is ``If''.

\subsubsection{Strategy Specification Matching}
\label{sec:pattern_matching}
For each strategy specification $s$ of a repair approach, \linebreak
\toolname checks if $s$ matches with the AST actions ($ES$) previously computed (line \ref{algo:matching} of \autoref{tool_algorithm}).
To this end, for each action $A_p$ specified in $s$, we check whether there exists an actual action in $ES$ that affects the nodes specified by $A_p$. The details of the matching process can be found in the study of Martinez et al. \cite{martinez2019coming}.

\revision{Note that \toolname considers a commit to be in the search space of repair approach r, only if all the changes in the commit are covered by a single strategy instance of r. We call such commits, \emph{single-instance} fixes. For example, a fix by GenProg that removes multiple statements from different methods is not considered as a repair-space commit by \toolname. On the other hand, a fix that removes a single statement that contains multiple-lines is indeed considered to be in the search space of GenProg by \toolname.
It is worth mentioning that \toolname already has the potential to detect multi-instance fixes as well. However, possible overlaps between multiple strategy instances in a single fix can lead to a high level of noise in \toolname detection algorithm. \revisiontwo{Since most existing repair tools create single-instance fixes in practice~\cite{durieux2019empirical}, we ignore multi-instance commits in the current version of \toolname.}}

\subsection{Final Checking}
\label{sec:post_matching}
In order to make sure that the source code changes from the identified commits lie in the search space of detected repair approaches, we also check particular rules that repair approaches follow for generating patches (line \ref{algo:post_checking} of \autoref{tool_algorithm}). We call these rules the \emph{post-matching rules}. These rules determine how a repair approach synthesizes new code.

The post-matching rules can be divided into two groups:
\begin{inparaenum}[\it 1)]
    \item rules specifying how the ingredients are extracted from the considered scope, and
    \item rules specifying how the ingredients are merged together to form new code fragments that are used in the patch.
\end{inparaenum}

As an example, Cardumen \cite{martinez2018ultra} considers all the variables and literals in the scope as repair ingredients. Next, it takes an existing \textit{expression} and replaces its variables and literals with extracted ingredients of the same type to make a new expression. This new expression is then used to generate a patch.

The commits given as the input of this step that follow the post-matching rules are considered as the detected \repairspacecommits.

\subsection{\revision{Meeting the Challenge}}

\begin{lstlisting}[language=XML, caption={\revision{Simple specification for GenProg strategies.}}, captionpos=b, label=lst:genprog_strategies, float=bt, floatplacement=bt, escapechar=|]
<pattern name="genprog_simple">
    <entity id="1" type="*" role="Statement"/>
    <action entityId="1" type="*"/>
</pattern>
\end{lstlisting}

\begin{lstlisting}[language=XML, caption={\revision{Second specification for GenProg strategies.}}, captionpos=b, label=lst:genprog_strategies2, float=bt, floatplacement=bt, escapechar=|]
<pattern name="genprog_indirect">
    <entity id="1" type="*" role="Statement"/>
    <entity id="2" type="*" role="*">
        <parent parentId="1" distance="100000"/>
    </entity>
    <action entityId="2" type="*"/>
</pattern>
\end{lstlisting}

\revision{As explained in \autoref{sec:challenges}, the modifications of an AST in a commit can be represented in various ways. The edit script that GumTree produces is only one of such representations. To make sure that we recognize any possible correspondence between an edit script and a repair strategy, we take two steps as follows.}

\revision{\textbf{1) Designing all relevant specifications:} We design all possible specifications whose matching edit scripts can be an instance of the target repair strategy. \revisiontwo{The \linebreak ``\texttt{get\_specifications}'' method at line \ref{algo:get_specs} of \autoref{tool_algorithm} retrieves all of these specifications}. For example, consider the repair strategies of GenProg. \autoref{lst:genprog_strategies} is the simple and natural specification that encodes those strategies. It represents any action (move/insertion/removal) on a statement. The code change in \autoref{lst:replace_statement_ex} is inserting a new statement (at line 7), therefore it is an instance of this simple specification for GenProg strategies. However, the edit script $ES$ that GumTree generates for this change is an ``operand removal'' action that removes ``\texttt{+val2}''. $ES$ does not match the simple specification, as the specification requires an action on a statement, while $ES$ contains an action on an operand.}

\revision{To make sure that \toolname does not miss strategy instances such as \autoref{lst:replace_statement_ex}, we also design and add the second specification shown in \autoref{lst:genprog_strategies2} for GenProg. This specification represents an action (move/insertion/removal) on an element inside a statement. This specification matches with the $ES$, since removing ``\texttt{+val2}'' from line 6 of \autoref{lst:replace_statement_ex} is counted as a removal of an operand inside the statement ``\texttt{var=val1+val2;}''. Therefore, by adding the specification in \autoref{lst:replace_statement_ex}, we also catch this strategy instance.}

\revision{\textbf{2) Filtering out non-instance matches:} By designing and considering all relevant specifications, we may match $ES$ with a specification for a repair strategy $rs$, while the code change represented by $ES$ is not an instance of $rs$. In our final checking step, we also check whether each matched edit scripts represents a change that is actually a strategy instance. 
This final check should be particularly tailored for each repair strategy. For example, for GenProg, we make sure the new statement is similar to an existing statement before repair.
}

\revision{Specifications designed in the first step match only a few types of commits and filter out the rest. 
This enables us to carefully design proper checks for the commits that go to the ``filtering out non-instance matches'' step. 
\revisiontwo{This whole process is a fast and accurate mechanism for matching edit scripts and specifications.}}

\subsection{Repair Approaches Considered}
\label{sec:repair_tool_considered}

\begin{table*}[!htb]
%\captionsetup{font=normalsize}
\footnotesize
\centering
\caption{Specification of the Search Space of 8 Notable Repair Approaches.} \label{tab:tools}
\begin{tabular}{| m{1.3cm} | m{7.6cm} | m{7.6cm} |}
\hline
\textbf{Name} & \textbf{Excerpt of strategy specifications} & \textbf{Excerpt of post-matching rules} \\ \hline \hline
Arja &
Removing a statement;

Inserting a new statement;

Replacing a statement with a new one. & 
The new statement should be a copy of an existing statement, while the variables, methods, and literals can be replaced by other variables, methods, and literals of the same type in the scope.
\\ \hline 

Cardumen &
Replacing an expression with a new expression. &
The new expression should be a copy of an existing expression, while the variables, and literals can be replaced by other variables, and literals of the same type in the scope.

Moreover, the new expression should have the same return type as the old one.
\\ \hline 

Elixir &
Replacing the declaration type of a variable with a wider one (e.g float to double);

Replacing a return expression with a new expression;

Moving a statement into a new if-statement. The condition of the new if-statement checks if one of the variables used in the statement is not null;

Mutating a binary operator. e.g., ``<'' to ``>'';

Replacing a method invocation with a new one;

Inserting a new method invocation;

Removing a predicate from the boolean expression of an if-condition or return statement;

Adding a predicate to the boolean expression of an if-condition or return statement;

Moving a statement into a new if-statement. The condition of the new if-statement checks if an array or collection access is in a range. &
All the code fragments used in a synthesized code should be collected from existing code. Specifically, a new method invocation should call a method that is already called in the scope. The argument list of a new method invocation should also be a list of literals and variables in the scope. For more details, please see the original paper by \cite{saha2017elixir}. 
\\ \hline 

GenProg &
Removing a statement;

Inserting a new statement;

Replacing a statement with a new one. &
The new statement should be exactly a copy of an exiting statement in the scope.
\\ \hline

jMutRepair &
Changing a unary or binary operator inside an if-condition. &
No post-matching rule.
\\ \hline

Kali &
Removing a statement;

Changing an if-condition to \textit{true}/\textit{false};

Inserting a return statement. &
No post-matching rule.
\\ \hline

Nopol &
Replacing an if-condition with a new if-condition;

Inserting a new if-statement and moving an existing statement into it. &
The new if-condition should consist of variables, methods, and literals that exist in the scope. \\ \hline 

NPEfix &
Moving a statement into a new if-statement. The if-condition checks if a variable used in the moved statement is not null;

Moving a statement into a new if-statement. The if-condition checks if a variable used in the moved statement is not null. Also, add an else-block which returns ``null'' or a new object or another variable of the desired type;

Moving a statement into a new if-statement. The if-condition checks if a variable used in the moved statement is not null. Also, add an else-block which executes the same statement but replaces the checked variable with a new object or another variable of the same type;

Inserting a new if-statement before a target statement. The if-condition checks if a variable used in the target statement is null. The corresponding then-statement sets the value of the checked variable to a new object or another variable of the same type. &
The variables used in synthesized code should be from existing code in the scope.
\\ \hline
\end{tabular} 
\end{table*}

\revision{In \toolname, we specify the search space of test-based repair approaches according to the following criteria.
First, they are included in Table 1 of Durieux et al. ~\cite{durieux2019empirical} study. Durieux et al.'s study provides a large list of repair tools that is frequently used by researchers as a reference for conducting empirical analysis of program repair \cite{qin2021impact,aleti2021apr,lin2020understanding}.
Second, the article presenting the repair approach provides us with enough information to specify the repair search space.
Third, the repair approach should have an explicitly defined search space so that we can specify it. For example, we exclude learning-based approaches, as their search space is hidden in the weights of their neural network.
According to those criteria, we select eight repair approaches: Arja \cite{yuan2018arja}, Cardumen~\cite{martinez2018ultra}, Elixir~\cite{saha2017elixir}, GenProg~\cite{le2011genprog}, jMutRepair~\cite{martinez2016astor}, Kali~\cite{qi2015analysis}, Nopol~\cite{xuan2016nopol}, and NPEfix~\cite{cornu2015npefix}.
\toolname tries to detect Java \repairspacecommits (the dataset is introduced in \autoref{sec:datasets}). Therefore, we consider the implementations of these approaches that repair Java programs. This means for GenProg and Kali, which have implementations for both Java and C, we consider their implementations for Java in jGenProg2 and jKali \cite{martinez2016astor}.}

In \autoref{tab:tools}, each row presents a brief overview of the strategy specifications and the post-matching rules that we consider to encode the search space of the corresponding repair approach. For instance, three strategy specifications are considered to encode the repair strategies employed by Arja, one for inserting a new statement, one for removing a statement, and one for replacing a statement. Moreover, in accordance with the process of synthesizing new statements in Arja, we have a post-matching rule: the new statement should be a copy of an existing statement, while the variables, literals, and methods can be replaced by other variables, literals, and methods in the scope.

\subsection{Implementation}
\label{sec:implementation}
We implement a prototype of our proposed method called \toolname. \toolname is built on top of Coming~\cite{martinez2019coming}. Coming is designed to mine instances of code change patterns in Git repositories. \toolname extends Coming by adding strategy specifications and post-matching rules for the considered repair approaches. The post-matching rules are implemented in Java and the strategy specifications are represented in the \textit{change pattern specification language} \cite{martinez2019coming} as noted in \autoref{sec:change_pattern_specification}. 
For all approaches based on code reuse, \toolname considers that the scope of ingredients is the \textit{same file} level. This means a \repairspacecommit must utilize ingredients from the same file as the repair location. \revisiontwo{Note that previous studies show that among human made patches that reuse existing ingredients, in 65\% of the cases all ingredients are selected from the exact same file \cite{yang2021were}. Therefore, we select the file scope as it helps us simplify the experiments without loss of generality.}
For sake of open science, \toolname is made publicly available \cite{RSCommitDetector}.

\section{Experimental Methodology}
\label{sec:methodology}
\subsection{Research Questions}

\newcommand\rqone{How do repair approaches compare to each other in terms of human-written patches that lie in their search spaces?}

\newcommand\rqtwo{To what extent do the search spaces of repair approaches overlap according to our lightweight analysis?}

\newcommand\rqthree{What is the recall of \toolname for \repairspacecommit detection? }

\newcommand\rqfour{What is the precision of \toolname for \repairspacecommit detection? }

\newcommand\rqfive{\revision{How complex are the commit matching criteria that encode the repair search space of program repair approaches?}}

In this paper, we study five research questions. The first two concern a deep study of the commit coverage of program repair approaches over human-written past commits.

\begin{itemize}
    \item RQ1: \rqone \\ To answer this research question, we use \toolname to detect \repairspacecommits for the considered repair approaches, and thereby compute their commit coverage over a large dataset of real-world commits. Moreover, we conduct a manual study to measure the prevalence of \repairspacecommits that are actually fixing bugs.
    \item RQ2: \rqtwo \\ It is known that some repair strategies are shared between repair approaches \cite{arXiv-1802.03365}. Therefore, one can expect to see shared commits between search spaces. In this experiment, we investigate the extent of this search space overlap.
\end{itemize}

\revision{The next two research questions measure the accuracy of the matching mechanisms and the search space specifications employed by \toolname.}

\begin{itemize}
   \item RQ3: \rqthree
   \item RQ4: \rqfour
\end{itemize}

\revision{The last research question studies the complexity of search space specification in \toolname.}

\revision{\begin{itemize}
   \item RQ5: \rqfive \ \toolname specifies the search space of eight existing repair approaches. To study the difficulty of encoding other repair approaches, we analyze the complexity of strategy specification and post-matching rules already implemented.
\end{itemize}}

\subsection{Datasets}
\label{sec:datasets}

\begin{table}[t]
\centering
\footnotesize
\begin{threeparttable}
\caption{\revision{Dataset features.}}
\label{tab:datasets}
\begin{tabular}{l r r r}
    \toprule
	\textbf{\projectsFromBears} &  & \textbf{\#Commits} \\
    All 72 repos     &  & 55,309 \\
    Repos with at least 1000 commits
     &  & 43,000 \\
    Repos with fewer than 1000 commits
     &  & 12,309 \\
    Ex: github.com/apache/pinot
     &  & 1,000 \\
    Ex: github.com/2018swecapstone/h2ms
     &  & 931 \\
	\toprule
	\textbf{\groundtruthpatches} & \textbf{\#Bugs} & \textbf{\#Patches} \\
	\cline{1-3} 
	All 5 projects & 160 & 729 \\
	jfreechart & 15 & 112 \\
	closure-compiler & 54 & 91 \\
	commons-lang & 25 & 165 \\
	commons-math & 56 & 319 \\
	joda-time & 10 & 42 \\
	\bottomrule
\end{tabular}
\end{threeparttable}
\end{table}

In this paper we use two datasets:
\begin{inparaenum}[\it 1)]
    \item a curated set of open-source repositories and their commits. This dataset is used to answer RQ1, RQ2, and RQ4.
    \item a set of patches (i.e., source code changes) that are generated by automatic program repair approaches. We employ this dataset to answer RQ3.
\end{inparaenum}
They are collected as follows.

The dataset of repositories, which we call \projectsFromBears, contains all projects that are included in the bug benchmark \textsc{Bears} \cite{madeiral2019bears}.
\revision{\textsc{Bears} contains bugs and their respective patches collected from 72 distinct open-source Java projects.}
We consider \textsc{Bears} since it has the largest number of projects among datasets of its type (72 versus 6 for Defects4J \cite{just2014defects4j}).

\revision{For each project in \projectsFromBears, we consider the last 1,000 commits as of October 08, 2021. For projects with less than 1,000 commits, we consider all commits. As shown in \autoref{tab:datasets}, the total number of commits considered in this dataset is $55,309$. 43,000 commits are selected from 43 projects with at least 1,000 commits and the remaining 12,309 commits are selected from \linebreak projects with fewer than 1000 commits. For example, ``pinot'' has 1,000 commits in \projectsFromBears, while ``h2ms'' has no more than 931  commits.}

\revision{It is to be noted that \projectsFromBears contains all types of commits. 
This is a deliberate decision because there is no accurate automatic oracle for determining bug-fix commits. Also, manual detection of bug-fix commits in a very large dataset of commits is practically impossible.}

The second dataset, named \groundtruthpatches, is a benchmark of patches produced by the eight automatic repair approaches that we consider in this work.
This dataset is built as a subset of the dataset DRR \cite{He2019DDR} and from the dataset NPEfix \cite{durieux2017dynamic}.
We choose DRR because it is the largest curated collection of patches automatically generated by repair approaches on the Defects4J dataset \cite{just2014defects4j}. 
It includes patches for all approaches we consider, except for NPEfix.
Note that DRR also includes patches generated by approaches that we do not consider, those patches are not included in \groundtruthpatches.
Second, the NPEfix patches come from the original study of Durieux et al.~\cite{durieux2017dynamic}, we use them because DRR does not contain patches for NPEfix while bug-fix commits related to null-pointers are important in practice.

% Note that we have two assumptions about the changes in \repairspacecommits:
% \begin{inparaenum}[\it 1)]
% \item the scope of ingredients used in those changes should be \textit{same file}, and
% \item only one change should be made by the repair approach.
% \end{inparaenum}
% There are ground-truth patches in DDR which do not follow our assumptions. For example, some GenProg ground-truth patches insert statements selected from source files other than the changing file. Which means in these ground-truth patches the scope of ingredients is not at the \textit{same file} level. By manually investigating the ground-truth patches, we identify the ground-truth patches that violate our assumptions.
% We do not include this set of patches in \groundtruthpatches.
\revision{In total, \groundtruthpatches contains $729$ patches. As presented in \autoref{tab:datasets}, these patches are generated for $160$ different bugs. The number of bugs and patches per project are also shown in \autoref{tab:datasets}. All five projects and their bugs are from the well-known Defects4J~\cite{just2014defects4j} dataset and the patches come from the carefully curated DRR dataset~\cite{He2019DDR} and NPEfix study~\cite{durieux2017dynamic}. For example, 15 bugs from ``jfreechart'' and their 112 fixing patches are considered.}

\subsection{Protocol for RQ1}
\label{sec:experiment_1}
RQ1: \rqone \\
We use \toolname to answer RQ1. \revision{We run \toolname on all commits collected in the \projectsFromBears{} dataset.}
\revision{This experiment is carried out in the form of a sequence of executions. Each execution is represented by a pair, like \textit{\textless a, r\textgreater}, where $a$ is a repair approach and $r$ is a repository.} In total, we perform 576 (8 approaches $\times$ 72 projects) executions. In each execution, \toolname goes through the commits of \textit{r} and checks if the source code changes in each commit are in the search space of \textit{a}. This experiment is conducted on a server with an Intel Core Processor running at 2299.996 MHz using 8GB of RAM, running Ubuntu version 18.04.

The result of this experiment is a set of \repairspacecommits for each pair of repair approach and repository. \revision{Based on those results, we compute the commit coverage for each repair approach to find which of them perform repairs that are similar to human-made changes in open source projects. \revisiontwo{In addition to calculating the percentage of all commits that are covered by the considered repair approaches, we compute the ratio of commits with changes in exactly one source file that lie in the search space of repair approaches. This gives use a better sense of the coverage of the considered repair approaches over the particular commits that we target in this experiment.}}

Next, we perform a comprehensive manual analysis to identify the detected \repairspacecommits that are bug-fixing. The process is as follows: 1) each detected \repairspacecommit is labelled by two persons, 2) the labels could be ``bug-fixing'', ``not-bug-fixing'', or ``dont-know'';
3) in case of a judgment conflict between the two persons, a third person annotates the commit to break the tie. Ten people participated in this annotation process, seven PhD students, one Postdoc researcher, and two professors, all working on areas close to automatic program repair. \revision{The results of this manual analysis tell us how many of the \repairspacecommits detected by \toolname are bug fixes.} The manual analysis process is extensively documented in our companion repository~\cite{RSCommitDetector}.

\subsection{Protocol for RQ2}
\label{sec:prot_rqtwo}
RQ2: \rqtwo\\
To answer RQ2, for each repair approach, we compute the number of its \repairspacecommits that also lie in the search space of another repair approach. For this purpose, we consider the \repairspacecommits that are detected in response to RQ1.

\revision{Moreover, we group the repair approaches by the type of changes they apply and analyze the overlap between the search space of these groups. This analysis reveals well-known types of changes that repair approaches apply and cannot be replaced by other types of changes. 
By considering the repair strategies listed in \autoref{tab:tools} and per authors' consensus, we categorize repair approaches into three groups. Three of the repair approaches make a change in if-conditions of the subject program (jMutRepair, Nopol, and NPEfix) and three of them add/remove/replace statements (Arja, GenProg, and Kali). We call these groups ``if-change'' and ``s-change'' (for statement change), respectively. We put the remaining two approaches (Cardumen and Elixir) in a third group and call that group ``other''. Note that Kali removes a functionality in program, which can be done either by removing a statement, inserting a return statement, or changing an if-condition to \texttt{true} or \texttt{false}. This means Kali is able to make a change on both statements and if-conditions. However, by manually reviewing our early results, almost all Kali \repairspacecommits remove a statement. Thus, we put it in the s-change group.} Finally, we compute the overlap of \repairspacecommits among these three groups.

If the results of this experiment show a repair approach has many unique commits, it means that its adds something useful and original capabilities to the state-of-the-art of automatic program repair.
% \revision{This is a good example of an analysis of the search spaces of repair approaches that is conducted based on the data provided by \toolname.}

\subsection{Protocol for RQ3}
\label{sec:recall_protocol}
RQ3: \rqthree\\

\revision{To answer RQ3, we design an experiment to determine the recall of \toolname, that is, how many of the patches actually generated by a repair approach are correctly detected by \toolname. For this, we run \toolname on the \groundtruthpatches dataset because we need a fully labelled dataset. For the same reason, it is impossible to use \projectsFromBears in RQ3 because the commits in \projectsFromBears are not labelled.}

Ideally, \toolname should be able to detect all of these source code changes as instances of corresponding repair approaches. However, it might happen that some ground-truth patches cannot be detected by \toolname due to the difficulty of encoding the search space of repair approaches. We call these patches the \textit{false negative patches (FN)}.

Finally, the recall of \toolname is calculated according to \autoref{eq:recall}. In this equation \textit{GT} represents the set of all patches in the \groundtruthpatches. \textit{TP} is the set of \textit{true positives}: the set of ground-truth patches that are detected as \repairspacecommits by \toolname. Note that $|TP|=|GT|-|FN|$ because \textit{TP} is the set of ground-truth patches that are not among the false negatives.

\begin{equation}
\label{eq:recall}
\mathit{recall}=\frac{|TP|}{|GT|}
\end{equation}

Since \toolname is the first tool of its type, there is no other work that we can directly compare against. However, a close tool is PPD (Patch Pattern Detector), which detects instances of repair patterns~\cite{madeiral:hal-01851813}. The repair patterns that PPD looks for are extracted from the code changes in Defects4J dataset and the tool is also evaluated on Defects4J. We compare the recall of our tool against PPD.

\subsection{Protocol for RQ4}
\label{sec:precision_protocol}

RQ4: \rqfour\\
We measure the precision of \toolname as follows: we randomly select a sample of $n$ \repairspacecommits for each tool in the \projectsFromBears dataset. Next, we carry out a manual analysis to decide if the detected \repairspacecommits actually lie in the search space of corresponding repair approaches or not.

We select a value of $n$ such that the overall manual work stay under two days over all analysts. Recall that the annotators have to be trained to be fully familiar with the corresponding repair approaches.
According to this criterion, we select $30$ commits per repair approach, each of them being analyzed by three analysts.

This manual analysis is made by seven analysts in total, all of whom are researchers in the field of automatic program repair: three PhD students, two postdoctoral researchers, and two professors. Each commit is annotated by two analysts. If the first two annotations conflict with each other, a third analyst annotates to break the tie. All results from this experiment are publicly available (see \cite{RSCommitDetector}).

\revision{The precision for each repair approach is computed per \autoref{eq:precision}. In this equation, \textit{true positive (TP)} represents the set of detected \repairspacecommits that are actual \repairspacecommits according to the manual investigation. Moreover, \textit{RSCommits} is the set of all considered \repairspacecommits for the current repair approach.}

\begin{equation}
\label{eq:precision}
\mathit{precision}=\frac{|TP|}{|RSCommits|}
\end{equation}

Similar to the recall, we also compare the precision of our tool with that of PPD \cite{madeiral:hal-01851813}.

\subsection{\revision{Protocol for RQ5}}
\label{sec:prot_rqfive}
\revision{RQ5:} \rqfive\\
\revision{To answer RQ5, we compute relevant metrics for strategy specifications and post-matching rules (see \autoref{sec:change_pattern_specification} and \autoref{sec:post_matching}).
First, for measuring the complexity of a repair approach encoding, we define 3 variables: 1) total number of specifications, 2) number of actions and 3) number of entities inside the specifications.
When a repair approach has more than one corresponding specifications, the number of actions/entities for that approach is the sum of the number of actions/entities in all the related specifications.
We consider the lines of code (LOC) to measure the complexity of post-matching rules.}

\section{Experimental Results}
\label{sec:results}

We now present our experimental results on the commit coverage of program repair approaches.

\subsection{RQ1: Commit Coverage per Repair Approach}
\label{sec:results_rq1}

\begin{table*}[t]
\centering
\begin{threeparttable}
\caption{\revision{\revisiontwo{RQ1: The presence of repair-space commits in 72 open-source projects.}}}
\label{tab:stats}
\begin{tabular}{l r r r r r r r}
    \toprule
	Approach & \#RSC\tnote{a} & \%CC\tnote{c} & \%TCC\tnote{c} & BF\tnote{d} & NBF\tnote{d} & DN\tnote{d} & Exec. Time \\
	
    \midrule
	Arja & 263 & 0.47 & 3.46 & 51\% & 38\% & 11\% & 0.76s \\ 
	Cardumen & 219 & 0.39 & 2.88 & 63\% & 29\% & 8\% & 0.33s \\ 
	Elixir & 369 & \textbf{0.66} & 4.86 & 67\% & 24\% & 9\% & 1.72s \\ 
	GenProg & 181 & 0.32 & 2.38 & 46\% & 42\% & 12\% & 0.77s \\ 
	jMutRepair & 7 & 0.01 & 0.09 & 86\% & 14\% & 0\% & 0.87s \\ 
	Kali & 117 & 0.21 & 1.54 & 31\% & 56\% & 13\% & 0.29s \\ 
	Nopol & 174 & 0.31 & 2.29 & 81\% & 13\% & 6\% & 1.34s \\ 
	NPEfix & 33 & 0.05 & 0.43 & 90\% & 3\% & 7\% & 0.46s \\
	\midrule
	All & 747\tnote{b} & 1.35 & 9.85 & 62\% & 29\% & 9\% & 0.81 \\
	\bottomrule
\end{tabular}
\begin{tablenotes}
    \item[a] \footnotesize RSC stands for ``repair-space commits''. This column shows how many of the $55,309$ commits that are analyzed against the search space of all tools are detected as \repairspacecommits of this approach.
    \item[b] $747$ commits are detected \repairspacecommits for at least one repair approach. Note that this is not the sum of numbers in this column.
    \item[c] \revisiontwo{\%CC is the commit coverage of the corresponding approach. Also, \%TCC shows the percentage of commits with changes in exactly one source file that lie in the search space of corresponding repair approach. The total number of these commits is 7,583. For example, Elixir covers $0.66\% \; (369/55,309)$ of all considered commits and $4.86\% \; (369/7,583)$ of commits with changes in exactly one source file.}
    \item[d] BF, NBF, and DN represent the number of \repairspacecommits labelled as ``bug-fix'', ``not-bug-fix'', and ``dont-know'', respectively.
\end{tablenotes}
\end{threeparttable}
\end{table*}

\autoref{tab:stats} shows the results of our first experiment. In this table, each row represents the data for one repair approach.
The ``\#RSC'' column shows the number all commits that \toolname detects as \repairspacecommits. For each approach, ``\%CC'' presents the commit coverage, which is equal to the percentage of all $55,309$ human-written commits that are considered to be \repairspacecommits for that approach. \revisiontwo{``\%TCC'' is the percentage of our targeted commits that a) have changes in exactly one source file, and  b) lie in the search space of a repair approach.}
``BF'', ``NBF'', and ``DN'' indicate the number and percentage of \repairspacecommits labelled as ``bug-fix'', ``not-bug-fix'', and ``dont-know'', respectively, per our manual analysis described in \autoref{sec:experiment_1}.
Finally, the ``Exec. time'' column represents how many seconds it takes on average for \toolname to check if a commit is in the search space of the corresponding approach. \revision{\revisiontwo{For example, $263$ of commits are detected to be in the search space of Arja, which means Arja covers $0.47\%$ of all commits and $3.46\%$ of commits with changes in exactly one source file.} Moreover, $51\%$ of Arja's \repairspacecommits are labelled as ``bug-fix'' commits by the annotators.}

% finding 1
In total, $747/55,309 \, (1.35\%)$ commits are detected as \repairspacecommits for, at least, one of the repair approaches. 
\revision{Among the considered repair approaches, the top two approaches in terms of the commit coverage are Arja and Elixir. Given the strategies used by these approaches, this confirms the results of Martinez et al. \cite{martinez2014fix} showing that in a significant number of commits all the new lines are copied from the previous versions of the same file.}

% focus on 62%
The majority ($62\%$) of the detected \repairspacecommits are labelled as ``bug-fix''. This is in line with the fact that the encoded repair strategies are indeed related to the activity of bug fixing.

% focus on 29%
Interestingly, there is also a notable portion of the \repairspacecommits ($29\%$) that are not considered as bug-fixing, yet can likely be generated by a repair approach.  
This suggests that the considered repair approaches can also be used for purposes other than bug-fixing.
We manually analyze them and identify that there are two common types of ``not-bug-fix'' \repairspacecommits: 
\begin{inparaenum}[\it 1)]
\item commits that only change logging outputs, and
\item commits that remove unused code.
\end{inparaenum}
For example, \autoref{lst:not_bug_fix} is a commit from the ``pippo'' project that removes an unused variable ``rtHandler''. Although this commit is not labelled as bug-fix, it could be produced by Arja, GenProg, and Kali. 

\revision{There are also $9\%$ of \repairspacecommits that the analysts could not determine if they were bug-fixing or not. These \repairspacecommits are labelled as ``dont-know''. Most of these commits are hard to label due to uninformative commit messages and changes in uncommented parts of the code. This confirms that assessing arbitrary commits from the field is an inherently difficult task \cite{martinez2015mining,le:reliability-patch-assess}. The number of unclassifiable commits is low, much lower than the classified ones (9\% versus 91\%).}

\begin{lstlisting}[float=tb, style=diff, caption={Example of not-bug-fix \repairspacecommits.}, captionpos=b, label=lst:not_bug_fix]
%\RHilight%- RouteHandler rtHandler = route.getRouteHandler();
  route.getRouteHandler().handle(req, res, this);
\end{lstlisting}

Now, let us discuss the execution speed of our lightweight analysis.
The average time spent to check if a commit is in the search space of a repair approach is $0.81$ second. Elixir has the slowest strategy specifications and post-matching rules, \toolname needs $1.72$ second on average to analyze if a commit is in the search space of Elixir.
On the contrary, the \toolname configured for Kali takes only $0.29$ second per commit. Those numbers indicate that \toolname can scale to large repositories: for instance, analyzing $3.3k$ commits (the maximum number of commits for $99\%$ of Java projects on Github) against the search space of all tools would take approximately $6$ hours, which is acceptable given that it is one-short computation task.

\begin{mdframed}\noindent
    \textbf{Answer to RQ1: \rqone} \\
    According to our analysis, $1.35\%$ ($747/55,309$) of commits from 72 projects of dataset \projectsFromBears are detected as being in the search space of at least one repair approach.
    This experiment shows that our novel method enables researchers and practitioners to evaluate the potential of repair approaches in terms of their commit coverage.
    To the best of our knowledge, we are the very first to use commit coverage to compare program repair strategies.
\end{mdframed}

\begin{table*}[t]
    \caption{\revision{The overlapping of the repair approaches. Each row presents the percentage of detected \repairspacecommits in the search space of the corresponding approach that also lie in the search space of the rest of the approaches. For instance, $64\%$ of the \repairspacecommits of Arja (row 1) also lie in the search space of GenProg (column 2). Numbers on the table diagonal represent the unique \repairspacecommits, which means the commits that only lie in the search space of the corresponding repair approach. For example, 12\% of the commits in the search space of Arja are unique, they do not lie in the search space of any other repair approach. The color indicates overlap, the darker the cell the more the overlap.}}
    \label{tab:tool-overlapping}
    \centering
    \small
    \begin{tabular}{@{}l| r r r | r r r | r r}
        \toprule
  & \multicolumn{3}{c|}{S-change} & \multicolumn{3}{c|}{If-change} & \multicolumn{2}{c}{Other} \\
  
  & \multicolumn{1}{c}{Arja} & \multicolumn{1}{c}{GenProg} & \multicolumn{1}{c|}{Kali} & \multicolumn{1}{c}{jMutRepair} & \multicolumn{1}{c}{Nopol} & \multicolumn{1}{c|}{NPEfix} & \multicolumn{1}{c}{Cardumen} & \multicolumn{1}{c}{Elixir}  \\
 
 \midrule
Arja & \textbf{12\% (33)} & \cca {64}\% (170) & \cca {42}\% (111) & \cca {0}\% (0) & \cca {3}\% (8) & \cca {0}\% (1) & \cca {20}\% (54) & \cca {32}\% (85) \\
GenProg & \cca {93}\% (170) & \textbf{2\% (4)} & \cca {60}\% (110) & \cca {0}\% (0) & \cca {3}\% (6) & \cca {0}\% (0) & \cca {14}\% (27) & \cca {20}\% (38) \\
Kali & \cca {94}\% (111) & \cca {94}\% (110) & \textbf{5\% (6)} & \cca {0}\% (0) & \cca {0}\% (0) & \cca {0}\% (0) & \cca {0}\% (0) & \cca {0}\% (1) \\
jMutRepair & \cca {0}\% (0) & \cca {0}\% (0) & \cca {0}\% (0) & \textbf{14\% (1)} & \cca {85}\% (6) & \cca {0}\% (0) & \cca {0}\% (0) & \cca {85}\% (6) \\
Nopol & \cca {4}\% (8) & \cca {3}\% (6) & \cca {0}\% (0) & \cca {3}\% (6) & \textbf{29\% (52)} & \cca {10}\% (19) & \cca {17}\% (30) & \cca {55}\% (96) \\
NPEfix & \cca {3}\% (1) & \cca {0}\% (0) & \cca {0}\% (0) & \cca {0}\% (0) & \cca {57}\% (19) & \textbf{42\% (14)} & \cca {0}\% (0) & \cca {57}\% (19) \\
Cardumen & \cca {24}\% (54) & \cca {12}\% (27) & \cca {0}\% (0) & \cca {0}\% (0) & \cca {13}\% (30) & \cca {0}\% (0) & \textbf{42\% (94)} & \cca {36}\% (80) \\
Elixir & \cca {23}\% (85) & \cca {10}\% (38) & \cca {0}\% (1) & \cca {1}\% (6) & \cca {26}\% (96) & \cca {5}\% (19) & \cca {21}\% (80) & \textbf{41\% (153)} \\
        \bottomrule
    \end{tabular}
\end{table*}

%%%%%%%%%%%%%% RQ2 Analysis of the criteria
\subsection{RQ2: Overlap Between Repair Approaches}
\autoref{tab:tool-overlapping} shows the proportion of overlapping \repairspacecommits between each pair of repair approaches. Each cell presents the percentage of detected \repairspacecommits in the search space of an approach that also lie in the search space of another approach. For instance, $64\%$ of the \repairspacecommits of Arja (row 1) also lie in the search space of GenProg (column 2). 
On the opposite side, $93\%$ of \repairspacecommits of GenProg (row 2) also lie in the search space of Arja (column 1).
In the cells on the diagonal, the cell content represents the number and percentage of unique \repairspacecommits of the corresponding approach. \revision{A unique \repairspacecommits is a commit that only lies in the search space of one single repair approach and is not covered by other approaches.}

% finding: high variation of shared
The results from \autoref{tab:tool-overlapping} show that the ratio of shared \repairspacecommits vary significantly among repair approaches ranging from 0\% to 94\%. 
% finding on uniqueness
The table also shows that there are repair approaches with significant ratio of unique \repairspacecommits, clearly higher than others.
\revision{For example, NPEfix Cardumen, and Elixir have more than $40\%$ unique \repairspacecommits, while this number is less than $10\%$ for GenProg and Kali. In total, $47\% \, (357/747)$ of the \repairspacecommits are unique and the remaining $53\% \, (390/747)$ lie in the overlapping parts of the search spaces.} Here, uniqueness is a proxy to value, the approaches with high percentage of unique \repairspacecommits cannot be replaced by any other approaches.

\begin{figure}
\centering
% source is file XXXX
\includegraphics[width=3in]{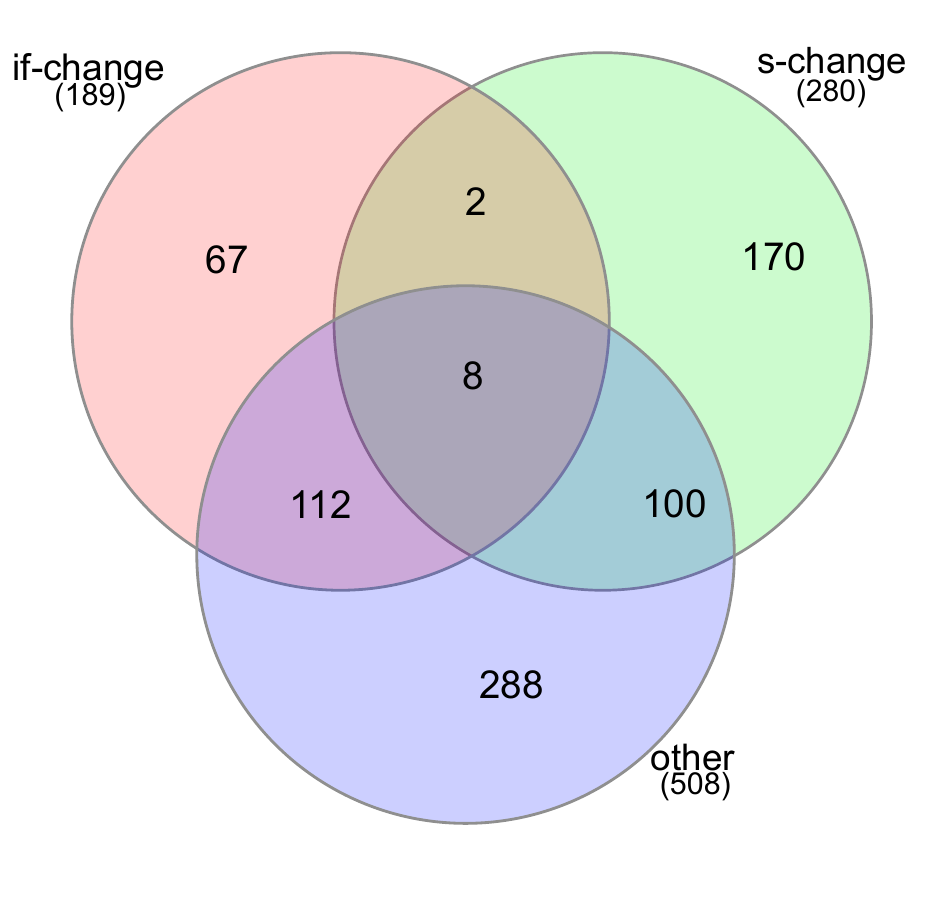}
\caption{Overlaps between the search space of groups of repair approaches.}
\label{fig:groups_overlap}
\end{figure}

% group analysis
As explained in \autoref{sec:prot_rqtwo}, we also divide the approaches into three groups (\textit{if-change}, \textit{s-change}, and \textit{other}) and compare their search spaces against each other. \autoref{fig:groups_overlap} depicts the overlaps between the search space of these groups of repair approaches. \revision{Note that the data represented in \autoref{fig:groups_overlap} cannot be fully retrieved from \autoref{tab:tool-overlapping}.} Recall that the search space of each group is the union of the search spaces of its members. The numbers on this figure show how many of the detected \repairspacecommits lie in the corresponding group. \revision{For example,
$170$ are exclusively in the search space of \textit{s-change} group and $112$ are common between \repairspacecommits of \textit{if-change} and \textit{other} but do not lie in the search space of \textit{s-change}. 
% seven commits are detected to lie in the search space of all groups, 
The overlap between if-change and s-change is small. There are 10 (=8+2) commits shared between \linebreak if-change and s-change, 8 of them are also shared with ``other'', while 2 of them are outside the search space of other. The total number of all \repairspacecommits in if-change and s-change is 459 (=67+2+8+170+100+112), therefore only 2\% (10/459) of their commits are common. The small overlap between if-change and s-change indicates that those two groups are complementary.}

\revision{Moreover, as illustrated in \autoref{tab:tool-overlapping}, the overlap between approaches of a group is usually higher than the overlap between approaches from different groups. For example, Kali has $111 \, (94\%)$ \repairspacecommits in common with other approaches from the s-change group, while it has no \repairspacecommit in common with approaches from if-change group.}
\linebreak These findings confirm that the grouping of approaches is meaningful.

\revision{\emph{Implication:} Based on the results from this experiment, we conclude that a general purpose repair approach should have at least three types of repair strategies: 1) statement replacement (s-change), 2) control-flow mutator (if-condition), and 3) expression replacement (Cardumen and Elixir). That is because no strict subset of these three categories covers the overall search space.}

\begin{mdframed}\noindent
    \textbf{Answer to RQ2: \rqtwo} \\
    \revision{Our results show that $47\% \, (357/747)$ of \repairspacecommits are unique to one specific repair approach, which is a novel result in the literature. NPEfix, Cardumen, and Elixir have the largest number of unique \repairspacecommits (more than $40\%$). 
    This little overlap shows that program repair research is producing approaches that are complementary in practice, and useful for practitioners in an integrated manner.}
\end{mdframed}

%% RQ3 recall
\subsection{RQ3: Recall of \toolname}
\label{section:results_recall}

\begin{table}[tb]
\centering
\caption{RQ3: Recall for each repair search space.}
\label{tab:recall}
\begin{tabular}{@{} l r r r @{}}
\toprule
	\textbf{Approach} & \textbf{\#GT} & \textbf{\#TP} & \textbf{Recall} \\
	\midrule
	Arja & 129 & 117 & 0.90 \\
	Cardumen & 129 & 118 & 0.91 \\
	Elixir & 37 & 31 & 0.83 \\
	GenProg & 116 & 101 & 0.87 \\
	jMutRepair & 52 & 52 & 1 \\
	Kali & 53 & 47 & 0.88 \\
	Nopol & 103 & 101 & 0.98 \\
	NPEfix & 110 & 107 & 0.97 \\
	\midrule
	Total & 729 & 674 & 0.92 \\
	\bottomrule
\end{tabular}
\end{table}

\autoref{tab:recall} studies the recall of \toolname.
Columns ``\#GT'' and ``\#TP'' indicate the number of ground-truth and true positive patches, respectively (see \autoref{sec:recall_protocol} for more details). The recall is computed according to \autoref{eq:recall}.

For instance, there are $129$ ground-truth patches for Arja and \toolname detects $117$ of them. Consequently, the recall for Arja is $0.90$.
\toolname has the lowest recall for Elixir ($0.83$) and the highest one for jMutRepair, a perfect recall of $1$.  \revision{The total recall is $0.92$, this is arguably a high recall, in line with the state-of-the-art related tool PPD \cite{madeiral:hal-01851813} (the total recall of PPD is $0.92$ as well).}

\autoref{lst:npefix_ex} is an example of a NPEfix ground-truth patch that is detected by \toolname. In contrast,
% missed caed
\autoref{lst:recall_ex} shows an example of a GenProg ground-truth patch that is not detected. In this example, GenProg replaces the statement in line 2 with the statement in line 3. However, \toolname finds that the only change is removing ``-index-1'' argument. Consequently, this patch is not detected as a GenProg repair-space patch. This shows that encoding the search space of a repair approach is fundamentally hard, and that a strategy specification may miss some cases.

\begin{lstlisting}[float=tb, style=diff, caption={Example of an undetected ground-truth patch.}, captionpos=b, label=lst:recall_ex]
  if (this.autoSort) {
%\RHilight%-   this.data.add(-index-1, new XYDataItem(x,y));
%\GHilight%+   this.data.add(new XYDataItem(x,y));
  }
\end{lstlisting}

\begin{mdframed}\noindent
    \textbf{Answer to RQ3: \rqthree} \\
    \revision{Out of $729$ ground-truth cases, we compute that the recall of \toolname is $0.92$, which is on par or higher than the closest related tools. Per repair approach, the recall has a minimum $0.83$ and a median of $0.90$, which is consistently high.} Therefore, we conclude that \toolname can be trusted in terms of detecting commits that actually lie in the search space of program repair approaches. Practitioners can rely on \toolname to compute commit coverage on their projects before doing the heavy-duty work of configuring and integrating the actual tool.
\end{mdframed}

%%%%%%%%%%%%%%%%%%%%
%% RQ4 PRECISION
\subsection{RQ4: Precision of \toolname}

\begin{table}[t]
\centering
\caption{RQ4: Precision for detected repair-space commits.}
\label{tab:precision}
\begin{tabular}{l r r r}
\toprule
	\textbf{Approach} & \textbf{\#RSCommit} & \textbf{\#TP} & \textbf{Precision} \\ 
	\midrule
	Arja & 30 & 29 & 0.96 \\
	Cardumen & 30 & 25 & 0.83 \\
	Elixir & 30 & 21 & 0.70 \\
	GenProg & 30 & 24 & 0.80 \\
	jMutRepair & 7 & 5 & 0.71 \\
	Kali & 30 & 29 & 0.96 \\
	Nopol & 30 & 18 & 0.60 \\
	NPEfix & 30 & 18 & 0.60 \\
	\midrule
	Total & 217 &  169 & 0.77 \\
	\bottomrule
\end{tabular}
\end{table}

The computed precision is reported in \autoref{tab:precision}. In this table, ``\#RSCommits'' and ``\#TP'' indicate the number of detected \repairspacecommits in the sample set and the number of true positives, respectively (see \autoref{sec:precision_protocol} for more details). The precision is computed due to \autoref{eq:precision}.

Recall that for each repair approach, 30 detected \repairspacecommits are randomly sampled and manually analyzed\footnote{Except for jMutRepair for which there are only $7$ repair-space commits in total.}. For instance, among the $30$ sampled detected \repairspacecommits for Arja, $29$ of them are manually marked as true positives, meaning there are actually potential Arja patches. Therefore, the precision for Arja is $0.96$.

We see that \toolname has the best precision for Arja and Kali, where only one single commit is wrongly detected as a \repairspacecommit. In total, $169$ out of $217$ sampled commits are true positives and the total precision is $77\%$. We observe that the total precision of \toolname is lower than the reported total precision of PPD ($91\%$)~\cite{madeiral:hal-01851813}, we explain that difference as follows: PPD was fine-tuned for Defects4J, while our experiment considers many more diverse commits and projects.

\autoref{lst:precision_tp} is an example of a true positive. This commit changes a ``=='' operator to a ``!='' operator and is correctly detected as a jMutRepair \repairspacecommit. On the other hand, \autoref{lst:precision_ex} presents an example of a false positive for Nopol. This commit changes the condition of an if statement which in theory is in the search space of Nopol. However, the Nopol manual analyst concluded that the new condition is too complex to be synthesized by Nopol, because Nopol does not support ternary expressions.

\begin{mdframed}\noindent
    \textbf{Answer to RQ4: \rqfour} \\
    Thanks to the careful design of the matching criteria, the precision of \toolname is $0.77$. It is never lower than $0.60$ for any of the considered repair approaches. This high precision is important for program repair research: future researchers can rely on \toolname to create specifically tailored benchmarks of commits corresponding to the search space of a given repair approach.
\end{mdframed}

\begin{lstlisting}[float=tb, style=diff, caption={Example of a correctly detected jMutRepair repair-space commits.}, captionpos=b, label=lst:precision_tp]
%\RHilight%- if ((union & 0x0800) == 0) {
%\GHilight%+ if ((union & 0x0800) != 0) {
    position.setLatitude(latitude);
    position.setLongitude(longitude);
\end{lstlisting}

\begin{lstlisting}[float=tb, style=diff, caption={Example of a wrongly detected Nopol \repairspacecommit. Nopol is not able to synthesize ternary expressions (conditions with ``?'' and ``:'' signs).}, captionpos=b, label=lst:precision_ex]
%\RHilight%- if (v1.equals(v2)) {
%\GHilight%+ if (v1 == null ? v2 == null : v1.equals(v2)) {
    return options.fn();
  }
\end{lstlisting}

%%%%%%%%%%%%%% RQ5 Analysis of the criteria
\subsection{\revision{RQ5: Complexity of Repair Search Space Encoding}}
\revision{The results of the experiment (\autoref{sec:prot_rqfive}) are shown in \autoref{tab:matchers_metrics}. Columns ``\#Specifications'', ``\#Actions'', and ``\#Entities'' indicate the total number of strategy specifications, actions and entities for each repair approach that is implemented in \toolname. \revisiontwo{As explained in \autoref{sec:change_pattern_specification}, a specification for a repair strategy outlines the modifications that the strategy performs on the AST nodes of a program. In this context, the modifications are called \emph{actions} and the AST nodes are called \emph{entities} (see \autoref{jmutrepair_pattern} as an example of a specification).
The ``LOC'' column of \autoref{tab:matchers_metrics} shows the number of Java code lines for the post-matching rules implementation.} For instance, three strategy specifications are designed to encode the repair strategies employed by Arja. These specifications consist of four actions and five entities in total; the post-matching rules for Arja are implemented in 343 lines of Java code.}

\revision{In total, we design $34$ strategy specifications with $51$ actions and $85$ entities to encode the search space of all systems. Among all the repair systems, Elixir search space has the most complex encoding specifications with $17$ actions and $28$ entities. The complexity of Elixir is a consequence of the large number of the different repair strategies that it adopts: for example it includes all ``expression update'', ``statement addition'', and ``wrap inside if-statement'' repair strategies. Moreover, the implementation of post-matching rules contain $1,806$ lines of code in total. Arja has the largest post-matching rules with $343$ lines. The complex post-matching rules for Arja result from the different techniques that it uses to generate a new statement: it can change any literal, variable, or even method of an old statement. This analysis shows that the complexity of specifications and post-matching rules grow as the number of strategies grows in a repair approach. }

\begin{table}[t]
\centering
\caption{\revision{RQ5: Features of Strategy Specifications and Post-matching Rules.}}
\label{tab:matchers_metrics}
\begin{tabular}{@{} l r r r r @{}}
\toprule
	\textbf{Approach} & \textbf{\#Specifications} & \textbf{\#Actions} & \textbf{\#Entities} & \textbf{LOC} \\ 
	\midrule
	
	Arja & 3 & 4 & 5 & \textbf{343} \\
	Cardumen & 3 & 5 & 5 & 273 \\
	Elixir & \textbf{12} & \textbf{17} & \textbf{28} & 288 \\
	GenProg & 2 & 2 & 3 & 176 \\
	jMutRepair & 2 & 4 & 6 & 78 \\
	Kali & 4 & 4 & 4 & 79 \\
	Nopol & 4 & 6 & 10 & 275 \\
	NPEfix & 4 & 9 & 24 & 294 \\
	\midrule
	Total & 34 & 51 & 85 & 1,806 \\
	\bottomrule
\end{tabular}
\end{table}

\revision{Finally, one may compare the difficulty of specifying the repair space and running the actual repair systems on past commits. Here is the analysis, with a subjective analysis in parentheses.
To run a repair system on past commits, one need:
1) that the system is publicly available (not always the case),
2) that the system can be executed on any commit (uncommon),
3) that the commit can be compiled (hard),
4) that the commit contains a test case (rare).
On the contrary, our approach only requires to design strategy specifications and post-matching rules, which is arguably a much more lightweight way of analyzing past commits against repair search spaces.}

\begin{mdframed}\noindent
    \textbf{\revision{Answer to RQ5:} \rqfive} \\
    \revision{We are able to encode the search space of eight repair approaches using $34$ strategy specifications and $1,806$ lines of code. Elixir is the hardest search space to encode, while Kali and jMutRepair are the easiest ones. 
    The biggest advantage of our approach to study the breadth of the search space of program repair is that it is purely static. Consequently, our approach can be considered as lightweight.}
\end{mdframed}

\subsection{Threats to Validity}
\label{sec:threats}

\textit{Complexity of search spaces:} Because of the complexity of code change analysis, there is no perfect encoding of repair search space. The encodings implemented in \toolname do not yield a perfect matching. There are different factors contributing to false positive and false negatives, incl. noise in the commit, suboptimality of the AST edit script, and corner-cases of the repair approaches not captured in the declarative search space specifications.

\textit{Tangled commits:} As explained in \autoref{sec:pattern_matching}, we consider a commit $c$ as a \repairspacecommit for approach $r$ only if all the changes in $c$ correspond to a repair strategy employed by $r$. However, it is known that repositories contain tangled commits where different changes are mixed in the same commit \cite{herzig2013impact}. By construction, tangled commits in which only a subset of the commit changes correspond to a repair strategy are not considered as \repairspacecommits. This contributes to under-estimating the proportion of \repairspacecommits.

\revision{\textit{Out-of-file ingredients:} As explained in \autoref{sec:implementation}, we consider the ingredients from the ``same file'' scope. This means the current version of \toolname does not detect \repairspacecommits that use ingredients outside the same file in their patch, such as some patches of jGenProg2/ASTOR when the tool is configured to use the package or application scope. \toolname provides the basic framework for considering other ingredient scopes. Researchers and practitioners willing to employ \linebreak \toolname can configure it to consider other scopes than the default one.}

\revisiontwo{\textit{Semantically matching commits:} \toolname considers a commit to be in the search space of an approach only if the changes of the commit are syntactically equivalent to a patch in the search space of that approach. However, there may be many other commits performing changes semantically equivalent to search space patches. By not matching these commits \toolname is an underestimation which is sound: practitioners can only be positively surprised about the actual capability of APR approaches.}

\revisiontwo{\textit{Dynamic steps of repair:} As shown in \autoref{tab:pipeline_eval}, program repair approaches consist of both static and dynamic steps. \linebreak \toolname is focused on those static steps of repair. This means \toolname is oblivious to the challenges related to the dynamic steps, such as fault localization. This indicates that the potential of repair approaches computed by \toolname is an idealization, and it is by construction higher than what a tool actually delivers.}

\section{Discussion}
\label{sec:discussion}

\begin{table}[t]
\centering
\small
\caption{\revision{Prevalence of different types of commits in the \projectsFromBears dataset.}}
\label{tab:datasets_dissection}
\begin{tabular}{ l r }
\toprule
	\textbf{Commit Type}  & \textbf{\#Commits} \\ 
	\midrule
	All commits & 55,309 \\
    \quad Commits with no source change & 28,887 (53\%) \\
    \quad Commits with source change & 26,422 (47\%) \\
    \quad \quad Changes in multiple source files & 18,839 (34\%) \\
    \quad \quad Changes in exactly one source file  & 7,583 (13\%) \\
    \quad \quad \quad Without strategy instance & 3,795 (7\%) \\
    \quad \quad \quad With strategy instance & 3,788 (6\%) \\
    \quad \quad \quad \quad Instance not fully covering the commit & 3,041 (5\%) \\
    \quad \quad \quad \quad \textbf{Repair-space commits} & \textbf{747 (1.35\%)} \\
	\bottomrule
\end{tabular}
\end{table}

\subsection{Actionable Implications}
\label{sec:usage}

% what people can do
Here, we summarize the uses of our proposed method for researchers and practitioners.

\revision{\textbf{Prototyping of New Repair Approaches by Researchers:} The design space of program repair is very large \cite{matiasjss} and there are many different potential strategies for generating code fragments and forming patches. \revisiontwo{There are also tools that create new repair strategies by mining and analyzing human-written patches \cite{koyuncu2018fixminer}. A number of repair strategies and code synthesis techniques are already implemented in repair tools, such as TBar \cite{liu2019tbar}, CapGen \cite{wen2018context}, SimFix \cite{jiang2018shaping}, etc. These tools let us assess the performance of the particular repair approaches they have implemented. However, there are still many repair strategies left and implementing them for real and assessing their effectiveness with full execution can be very time consuming.} Our method and tool can be used by program repair researchers to assess the search space they envision without taking the hard path of fully implementing a tool. More specifically, \toolname enables researchers to measure the extent to which human-written commits are covered by the envisioned search space without full execution. \revisiontwo{It is worth noting that, as described in \autoref{sec:eval_types}, \toolname concentrates on the static steps of repair and provides a lightweight method to analyze repair search spaces. To have a comprehensive evaluation of the actual patches delivered by repair tools, the dynamic steps of repair should also be taken into account.}}

\begin{mdframed}\noindent
    \revision{Guideline for researchers: Use \toolname to identify promising repair strategies and filter out those that do not cover many real-world commits.}
\end{mdframed}

\revision{\textbf{Evaluation of the Potential Value of Using Program Repair by Practitioners:} The practitioners who consider automatic repair need to first assess the potential of repair tools on their own projects. 
Execution of existing tools is an option, but there are two major obstacles to execute them on real world projects. \revisiontwo{First, configuring a repair tool and actually executing it on a diverse set of projects is hard \cite{kechagia2021evaluating}. For example, TBar \cite{liu2019tbar} is a template-based repair tool whose current version is designed to run experiments only on the Defects4J dataset. It takes extensive work to configure TBar such that it can be run on any arbitrary project \cite{Applelid1612596}.} Second, many past bug-fix commits do not come with a failing test case \cite{madeiral2019bears}.}

\revision{Our proposed method can help practitioners assess the potential value of repair tools without facing the mentioned obstacles of full execution. Our contribution allows practitioners to quickly measure how many of their historical changes lie in the search space of repair approaches. This can be done without requiring heavy changes in the target project, configuring of a repair tool or having failing test cases for all past commits. This gives developers an estimation of their own bug-fixes that are in the search space of certain repair tools.}

\begin{mdframed}\noindent
    \revision{Guideline for practitioners: Use \toolname to quickly discard repair tools whose search spaces do not cover many past commits of their software project under consideration.}
\end{mdframed}

\begin{lstlisting}[float=t, style=diff, caption={Commit \revision{ca4f6aac} in ClassGraph, which renames ``head'' variable to ``tempFile''.}, captionpos=b, label=lst:refactoring_ex]
%\RHilight%+ final File head = tempFiles.remove();
%\RHilight%+ final String path = head.getPath();
%\RHilight%+ final boolean success = head.delete();
%\GHilight%+ final File tempFile = tempFiles.removeLast();
%\GHilight%+ final String path = tempFile.getPath();
%\GHilight%+ final boolean success = tempFile.delete();
\end{lstlisting}

\subsection{\revision{Dissection of Non Repair-space Commits}}

\revision{Recall that in our analysis, we consider all types of commits from a repository.
We showed in \autoref{sec:results_rq1} that a small percentage of commits ($1.35\%$) are \repairspacecommits.
Now, we conduct a study on our \projectsFromBears dataset (see \autoref{tab:datasets}) to find out why so many commits are not \repairspacecommits.}

\revision{As shown in \autoref{tab:datasets_dissection}, 53\% (28,887/55,309) of the commits do not contain changes in source files of the application (this also includes commits that only modify code related to test cases). Since all of our considered repair approaches are focused on fixing the source code, these commits lie outside our considered repair space. We note that this is promising for program repair which fixes build configuration files, e.g. \textsc{BuildMedic}~\cite{macho2018automatically}.}

\revision{Among the 26,422 commits that contain source code \linebreak changes, 18,839 commits (=34\% of all 55,309 commits) contain changes in multiple files. These commits are not considered as \repairspacecommits, since as described in \autoref{sec:input}, \toolname targets commits whose changes are completely covered by a single strategy instance. This leaves us with 7,583 commits whose changes are concentrated in a single Java source file. Out of these 7,583 commits, 3,795 (=7\% of 55,309) commits do not contain any strategy instances.}

\revision{Per our manual analysis on a random sample of 30 commits without strategy instances, they can be divided into three common groups. First, there are commits that only change documentation (i.e., code comments).
Second, there are commits that introduce novel code fragments that can not be generated by any existing repair approach (ex., a new string not used in the program before). 
Third, there are commits that change parts of the code that cannot be changed by our considered repair strategies (ex., changes in method/class names).
Each of these three types of changes can be subject to automation in future repair approaches.}

\revision{Finally, among the 3,788 commits with changes in a single source file that contain at least one strategy instance, 3,041 (=5\% of 55,309) commits have changes that are not fully covered by the detected strategy instance. This result is promising as it shows that many of the commits can at least be partially synthesized \cite{beyer2021towards}.}

\revision{By excluding all the commits that do not lie in the search space of any considered repair approaches, we end up with 747 (1.35\% of 55,309) commits that are \repairspacecommits. These are the commits that can be safely used to study the characteristics of repair search spaces.}

\subsection{\revision{Bug Type Distribution in the Wild}}
\revision{Researchers have studied the distribution of various bug \linebreak types in real-world projects and proposed automatic techniques for classifying bugs \cite{thung2012automatic,catolino2019not}. Understanding the common types of bugs can give us a better insight into developers mistakes that lead to defects in programs. An interesting point of discussion is how \repairspacecommit detection relates to assessing the prevalence of bug types in the wild.}

\revision{\toolname looks for correspondence between repair strategies and commits in the wild. Since the repair strategies are designed to repair buggy programs, most of the commits corresponding to them (i.e, \repairspacecommits) are indeed bug-fix commits, as shown by our results for RQ1 (\autoref{sec:results_rq1}) demonstrating that 62\% of \repairspacecommits are indeed fixing bugs.
A deeper look at them reveals that these bug-fix commits can be categorized into different ``bug types'' based on their corresponding repair strategies. For example, the repair strategies employed by NPEfix are designed to avoid a \texttt{NullPointerException} in the code, which means its corresponding bug type might be a missing null check.}

\revision{In a sense, our proposed method detects those \textit{types of bug-fix commits} for which there exists an APR repair strategy. The scope of considered bug types is closed.
In the contrary, in dedicated studies like \cite{thung2012automatic} and \cite{catolino2019not}, the scope is open, they analyze all possible bug types, and not only those for which an APR repair strategy exists. As future work, \toolname can potentially be extended to label the commits with the bug types they contain, similarly to ADD \cite{madeiral:hal-01851813}.}

%%%%%%%%%%%%%%%%%%%%%%%%%%%%%%%%%%%%%%%%%%%%%%%%%%
\section{Related Work}
\label{sec:related}
\subsection{Analysis of the Redundancy Assumption}

The key assumption behind GenProg is that the patch reuses some code from elsewhere in the program, this is called the redundancy assumption.
Previous works have investigated this assumption.
Barr et al. \cite{Barr2014Surgery} and Martinez et al. \cite{martinez2014fix} studied the assumption behind GenProg \cite{le2011genprog}:  patches are synthesized using fragments of code already written in the program under repair.
Those works measured the redundancy of a commit: for each commit, the redundancy is the percentage of code introduced that was already introduced by a previous commit.
Our approach is different, we verify that a single commit lies in the search space of a repair approach. Note that our post-matching rules also verify the redundancy of the introduced code for the repair actions that are based on the redundancy assumption. For example, the post-matching rule of GenProg verifies whether the statements included in a patch already exist in the buggy program.

\subsection{Mining Bug-fix Patterns from Bug Datasets}
\label{sec:related_mining_bug_fix}
Sobreira et al. \cite{defects4J-dissection} manually analyzed 395 ground-truth patches of Defects4J \cite{just2014defects4j} buggy programs.
They first identified abstractions, called repair patterns, occurring recurrently in patches and involving compositions of repair actions. 
They identified nine repair patterns from the patches in Defects4J, which span 373 patches of the dataset (94.43\%).

Madeiral et al. \cite{madeiral:hal-01851813} presented PPD, a detector of repair patterns in patches.
PPD performs source code change analysis at abstract syntax tree level and is able to detect the patterns found in Defects4J.
PPD and our work have important differences.
\begin{inparaenum}
\item[] First, they focus on a repair patterns that capture human-made changes, while we focus on repair strategies that characterize automated fixes from program repair approaches.
\item[] Second, our approach checks post-matching rules that are specific to repair approaches (as explained in Section \ref{sec:post_matching}), while PPD exclusively focuses on analyzing AST changes. 
\end{inparaenum}

\subsection{Mining Instances of Code Changes}
\label{sec:related_mining_code_change_instance}
There are different works that inspect bug-fix commits and patches with the goal of characterizing the bug-fixing activity.

Pan et al. \cite{pan2009toward} built a catalog with 27 bug-fix patterns that they manually identified by inspecting the history of seven open-source Java projects.
Then, they built a tool for detecting instances of such bug-fix patterns.
They finally reported the frequency of each bug-fix pattern.

Other works have mined Pan's pattern instances from other datasets.
Campos et al. \cite{Campos2017common} measured the prevalence of the five most common bug-fix patterns from \cite{pan2009toward}. 
For this purpose, they queried the Boa dataset~\cite{dyer2015boa} to find how many of the $4,590,405$ included commits follow each pattern. 
Islam et al. \cite{Islam2020bugfixpattern} have mined instances of 21 Pan's pattern from bug-fix commits done on 5 Java systems.

Those works have a different goal than ours.
First, they focus on mining instances of change patterns inside commits, while we focus on detecting \repairspacecommits.
Secondly,  they only do AST differencing, while we note
that AST analysis is insufficient to detect \repairspacecommits. As we presented in section \ref{sec:post_matching}, there are important additional rules that must be verified in order to confirm that a patch can be synthesized by a repair approach.

\subsection{Data-driven Program Repair}

Similar to Pan et al. \cite{pan2009toward}, Kim et al. \cite{kim2013automatic} manually inspected patches of open-source projects and from that inspection they defined 10 fix templates.
Then, they proposed Pattern-based Automatic Program Repair (PAR), a technique that applies these fix templates on faulty programs.
Other works have analyzed the presence of PAR's fix templates on bug-fix patches.
For example, Soto et al. \cite{Soto2016DeeperBugfixes} detected instances of PAR templates \cite{kim2013automatic} from bug-fixes done in Java projects.
For that, they analyzed 4,590,679 bug-fixing revisions queried from the Boa platform \cite{dyer2015boa}.
They found that the most frequent PAR template was "add or remove a branch condition" pattern which appeared in 4.23\% of the bug-fixing revisions. We discuss the differences at the end of this subsection.

Martinez and Monperrus \cite{martinez2015mining} built a probabilistic model of repair actions for guiding the navigation of the search space. \revision{For this purpose, they first compute the frequency of particular repair operators (ex., ``statement insert of method invocation'') in a large dataset of 89,993 real-world commits. This work is different from our study because Martinez and Monperrus check the commits against specific repair operators, while \toolname checks them against search space specifications that include repair strategies (comprising several operators) and post-matching rules.}

Soto and Le Goues \cite{Soto2018Predict} also created a probabilistic model of edit distributions that was used by a repair system to repair faster.
For that, the authors mined repair operators from bug-fixes done on the 500 most-starred Java projects on Github.
They encoded 19 operators in total, selected from those defined by GenProg \cite{le2011genprog}, PAR~\cite{kim2013automatic}, SPR~\cite{Long2015SPR} and three additional PAR templates. 

Ghanbari et al. \cite{Ghanbari2019Practical} have mined real bug-fix patches from the HDRepair dataset \cite{Le2016HDRepair} to measure the frequency of their repair operators implemented in their approach PraPR. 
Their goal was to further confirm the generality of the 18 PraPR mutators.
The  PraPR's ``\emph{MR mutator}'', which mutates method invocation instructions, is the most frequent operator: it appeared in 8.76\% of the bug-fix patches for the HDRepair dataset.

Those works and ours do the identification of \emph{instances} of bug-fix patterns. However, none of them identifies \repairspacecommits.
For that, our approach does advanced detection of strategy instances, and also checks rules that are specific to each repair approach. 
Moreover, none of those papers evaluates the accuracy and precision of their tool as we do in this paper.

\subsection{Analysis of the Patch Search Space}
\label{sec:rw:space}

Weimer et al. \cite{Weimer2013AE} presented AE, a repair approach that is specifically designed for optimizing the search space, using a cost model and multiple optimizations. For the evaluation of AE, the authors measured the size of the search spaces of AE and GenProg~\cite{LeGoues2012}. 
Our analysis is different, we do not measure the size of search spaces, we measure the inclusion of real past commits in those search spaces.

Long and Rinard \cite{long2016space} 
presented a systematic analysis of the SPR~\cite{Long2015SPR} and Prophet~\cite{long2016automatic} patch search spaces. 
With respect to our paper, the most related contribution of \cite{long2016space} is that they analyze the density of correct and plausible patches in the search space, and they characterize a trade-off between the size and sophistication of the search space. 
Our approach has a different goal, we do not analyze plausibility, we analyze past commits from repositories to assess applicability of program repair. \revision{More importantly, Long and Rinard run the repair tools to produce the candidate patches in their search space and perform further analysis on them, while we statically specify the repair search space to gain an understanding of their candidate patches.}

Petke et al. \cite{Petke2019GISpace} have surveyed the literature on the search spaces of genetic improvement, where they consider that program repair is one subset of such search spaces. Our paper provides a novel methodology for studying repair search space, which encodes search spaces with strategy specifications and rules, and it would be helpful for genetic improvement research beyond program repair.

\section{Conclusion}
\label{sec:conclusion}
% what we have done
\revision{In this paper, we have presented an original method for evaluating the breadth of the search space of program repair approaches by analyzing past commits. The key advantage of our approach is that it does not require to configure and execute repair tools on every single commit.
Using our approach, we analyze $55,309$ human-written commits from 72 Github projects.
Our original experiments validate the concept of using static analysis in order to study the breadth of the search space of program repair approaches.}

\section*{Acknowledgments}
This work was supported by the Wallenberg AI, Autonomous Systems and Software Program (WASP) funded by the Knut and Alice Wallenberg Foundation.

\bibliographystyle{elsarticle-num}
\bibliography{refs}

\end{document}